\documentclass[ reprint,amsmath,amssymb,aps,prl,superscriptaddress]{revtex4-1}
\usepackage{xparse}
\usepackage{hyperref}
\usepackage{bbm}
\usepackage{xr}
\usepackage{braket}
\usepackage{appendix}
\usepackage{graphicx}
\usepackage{dcolumn}
\usepackage{bm}
\usepackage{xcolor}
\usepackage[normalem]{ulem}

\usepackage{accents}

\usepackage{bbold}

\newcommand{\beginsupplement}{
\clearpage
\pagebreak
\setcounter{equation}{0}
\setcounter{figure}{0}
\setcounter{table}{0}
\setcounter{page}{1}
\makeatletter
\renewcommand{\theequation}{S\arabic{equation}}
\renewcommand{\thefigure}{S\arabic{figure}}
\onecolumngrid
\renewcommand{\thesection}{S\arabic{section}}
\section*{\large{Supplemental Material}}
}

\hypersetup{colorlinks=true,linkcolor=blue, anchorcolor = blue,citecolor = blue,filecolor = blue, urlcolor = blue}

\begin{document}

\title{Post-Quantum Quench Growth of Renyi Entropies in Low Dimensional Continuum Bosonic Systems}
\author{Sara Murciano}
 \email{smurcian@sissa.it}
\affiliation{SISSA and INFN, via Bonomea 265, 34136 Trieste, Italy}
 \author{Pasquale Calabrese}
 \email{calabrese@sissa.it}
\affiliation{SISSA and INFN, via Bonomea 265, 34136 Trieste, Italy}
\affiliation{International Centre for Theoretical Physics (ICTP), Strada Costiera 11, 34151 Trieste, Italy}
\author{Robert M. Konik}
\email{rmk@bnl.gov}
\affiliation{Condensed Matter Physics \& Materials Science Division,
 Brookhaven National Laboratory, Upton, NY 11973-5000, USA}
\date{
    \today
}

\date{\today}
\begin{abstract}
{The growth of Renyi entropies after the injection of energy into a correlated system provides a window upon the dynamics of its entanglement properties.  We develop here a simulation scheme by which this growth can be determined in Luttinger liquids systems with arbitrary interactions, even those introducing gaps into the liquid.  
We apply this scheme to an experimentally relevant quench in the sine-Gordon field theory.
While for short times
we provide an analytic expressions for the  growth of the second and third Renyi entropy,  to access longer times, we combine our scheme with truncated spectrum methods.}
\end{abstract}

\maketitle

\noindent\textbf{Introduction:} 

The time evolution of the Renyi entanglement entropy in out-of-equilibrium quantum field theory (QFT) nowadays plays a crucial role in disparate situations ranging from quantum gravity and black hole physics \cite{almheiri2020,stanford2022} to experiments in cold-atom and ion-trap setups \cite{2016,m-2018,vitale2021symmetryresolved,neven2021symmetryresolved}.
Very effective numerical techniques, based, e.g., on tensor networks allow us to compute their behaviour at not-too-long time scales in lattice systems \cite{SCHOLLWOCK201196,vidal2007,Daley_2004,white2004}. Conversely, simulation algorithms performing well for generic interacting field theories are not available yet (although continuous matrix product states \cite{cirac2010,jutho2013,Tilloy2019} represent a promising framework). The main goal of this work is to introduce and develop a new simulation scheme which will work for a large class of one-dimensional (1D) QFTs. %
The key idea is to use as a computational basis the one of the Luttinger liquid and write a general exact expansion for the Renyi entropies. The coefficients entering in such an expansion can be effectively calculated by truncated spectrum methodologies (TSM) \cite{yurov1990truncated,yurov1991truncated,James_2018}. 

The root of the effectiveness of our algorithm is that Luttinger liquids are a cornerstone for the description of a wide variety of quasi-1D systems \cite{Voit_1995,tsvelik_2003,Giamarchi}, spin-charge separation in 1D metals and nanotubes \cite{Laroche631,Wang2020}, power-law correlations of the dynamic structure function in 1D cold atomic systems \cite{CC1,atom_chip}, the fractionalization of magnons into spinons in quasi-1D spin chains \cite{Lake2009,Hirobe2016}, and even two-dimensional topological phases modeled through coupled wire constructions \cite{PhysRevB.89.085101}. 
Even when a Luttinger liquid is gapped out by an interaction, the underlying bosonic description of the unperturbed liquid provides an excellent starting point to understanding any underlying phenomena. 
Here our fundamental idea is using the unperturbed liquid as the starting point for the description of a wide variety of non-equilibrium dynamics. 

As a playground to  show the power and the potential of our approach, we focus on the problem of joining two Luttinger liquids. Such a protocol has been experimentally realized in \cite{schmiedmayer2018,Schmiedmayer2017,Schmiedmayer2018b} and has been theoretically discussed in many papers \cite{PhysRevLett.110.090404,PhysRevLett.121.110402,annurev-conmatphys-031214-014548,PhysRevA.100.013613,PhysRevLett.120.173601,essler2020,essler2021,Ruggiero_2021,Ruggiero_20212} (building upon previous work on non-interacting Luttinger liquids out-of-equilibrium \cite{Cazalilla_2006,PhysRevA.80.063619,PhysRevLett.112.246401,PhysRevLett.107.150602,PhysRevB.88.115144,Mitra_review}), but the time-evolution of the (Renyi) entanglement entropy is not tractable by other means.

The main object of interest here is the Renyi entropy 
\begin{equation}
    R_n(t) = (1-n)^{-1}\log {\rm Tr}(\rho(t)^n),
    \label{Rn}
\end{equation}
of a bosonic system with a time-dependent reduced density matrix, $\rho(t)$. 
We must mention that beyond the out-of-equilibrium scenario considered here, the Renyi entropies are of interest in several branches of physics.
For the condensed matter community, they provide a means to detect phase transitions and provide universal information on the nature of nearby critical points \cite{ent_rev1,ent_rev2,ent_rev3,ent_rev4}.  
For the high energy community, Renyi entropies play a key role in understanding holographic conformal field theories where they can be interpreted geometrically as the area of a dual cosmic brane \cite{Dong2016}, generalizing the famous Ryu-Takayanagi holographic formula \cite{RT} for the entanglement entropy in an AdS/CFT setting. 

While here we apply our machinery to the computation of time-dependent Renyi entropies, our framework also allows the determination of time-dependent relative Renyi entropies \cite{Ohya_Denes_2004,Araki:1976zv,RevModPhys.74.197,Audenaert2005}.  The relative entropy
\begin{equation}
    R_n(\rho(t)||\rho(0)) = -\partial_n({\rm Tr}(\rho(t)\rho(0)^{n-1})/{\rm Tr}(\rho(t)^n),
\end{equation}
can be viewed as a measure of the distinguishability of the time-evolved reduced density matrix from its $t=0$ value.  The relative entropy is not only a UV finite quantity, it is also closely connected to the entanglement spectrum of a system, a quantity which can be deeply connected to a system's topology \cite{LiHaldane}.

We stress once again that, although we present new and interesting results for the out-of-equilibrium Renyi entropies, the goal of this Letter is not providing fundamental physical insights on the specific quench dynamics of the coupled Luttinger liquids per se, but rather using it as a playground for a simulation scheme ideal to compute the entanglement in more generic 1D QFTs.


\noindent\textbf{Model for Non-Equilibrium Luttinger Liquids:}
To set the scene for our exploration of $R_n(t)$ in non-equilibrium Luttinger liquids, we will consider a canonical Hamiltonian density describing their dynamics:
\begin{equation}\label{ei}
    H(t)=\int^L_0 dx \frac{v_F}{8\pi}(\partial_x\phi^2 + \Pi^2) + 2J_1(t)\cos(\beta\phi).
\end{equation}
$\phi(x,t)$ is a real compact Bose field which admits the following mode expansion: 
\begin{equation}\label{eii}
\phi(x,t=0) =  \phi_0 + i \sum_{k\neq 0} \frac{1}{k}(a_k e^{\frac{i2\pi k x}{L}} - \bar a_{-k} e^{i\frac{i2 \pi k x}{L}}),
\end{equation}
where the $a_{-k}$'s are the bosonic creation operators for the oscillator modes and $L$ is the total length of the periodic system.  
The parameter $\beta$ is related to the Luttinger parameter, $K$, of the theory via $\beta=(2K)^{-1/2}$. $K$ determines the power law correlations in the model when $J_1=0$.

In order to explain how time-dependent Renyi entropies, $R_n(t)$, can be computed in a non-equilibrium setting, we need to review the Hilbert space of the $J_1=0$ theory, which here will serve as a computational basis for both our perturbation theory and numerics.  All states $|\Psi_i\rangle$ of the theory have the (unnormalized) form:
\begin{equation}\label{eiii}
|\Psi_i\rangle = \prod^{N_i}_{k=1} a_{-n_k}\prod^{\bar N_i}_{k=1} \bar a_{-\bar n_k}|\nu_i\rangle, ~~~ |\nu_i\rangle \equiv e^{i\nu_i \phi_0}|0\rangle.
\end{equation}
Here the $|\nu_i\rangle=n_i\beta$, with $n_i$ an integer, are plane waves states of the zero mode $\phi_0$ of the boson and $N_i$/$\bar{N}_i$ is the number of chiral/anti-chiral modes in the state $|\Psi_i\rangle$.

We now want to imagine that we have done a quantum quench or that $J(t)$ has a step-jump time dependence (more complicated time dependencies can be easily handled \cite{Hdsgi2020}).  We are going to suppose that we are tracking the time dependence of the state, $|\Phi(t)\rangle$, of the system via the following representation:
\begin{equation}\label{eiv}
|\Phi(t)\rangle = \sum_i \alpha_i(t)|\Psi_i\rangle,
\end{equation}
where $|\Psi_i\rangle$ are the states just discussed of the unperturbed bosonic theory.  Our focus on using the states of the unperturbed Luttinger liquids to describe entanglement dynamics differs from the form factor bootstrap approach where the emphasis is on the basis of gapped states of the sine-Gordon model \cite{oca,horvath2021branch}.  The corresponding density matrix of the system is
\begin{equation}\label{den_mat}
\rho (t) = \sum_{i,j}\alpha_i(t)\alpha^*_j(t) |\Psi_i\rangle\langle \Psi_j|.
\end{equation}
It will be with the density matrix in this form that we attack the problem of computing $R_n(t)$. 

\noindent\textbf{Time-Dependent Renyi Entropies:} 
Let us focus on the second Renyi entropy, $R_2(t)$, for simplicity.  Imagine that we perform a partial trace of region $B$ of the system ($=A\cup B$) from the density matrix in Eqn.~\ref{den_mat}. The second Renyi entropy will then take the form
\begin{eqnarray}\label{re_td}
R_2(t) &=& -\log\big(\sum_{i,j,i',j'}\alpha_i(t)\alpha_j(t)^*\alpha_{i'}(t)\alpha_{j'}(t)^* R_{i,j;i',j'}\big)\cr\cr
R_{i,j;i',j'} &=& {\rm Tr}_A ({\rm Tr}_B |\Psi_{i}\rangle\langle \Psi_{j}|{\rm Tr}_B |\Psi_{i'}\rangle\langle \Psi_{j'}|).
\end{eqnarray}
The object $R_{i,j;i',j'}$ is different than that normally considered.  If all the $|\Psi_i\rangle$'s are the same and are relatively simple (i.e. primary) states, we recover an object first studied in Ref. \cite{Sierra1,Sierra2} where the Renyi entropies of excited states in a conformal field theory were considered.  In the case when $i=i'$ and $j=j'$, the quantity at hand is related to the relative entropy, something that has been studied for the case of bosonic theories \cite{PhysRevLett.113.051602,rc,mrc_2019}.  The most general case $i\neq i'\neq j \neq j'$ has only been considered for low-lying descendant states in free fermionic theories \cite{Palmai,Palmai1}.  Here we exploit our recent development of general closed form expressions for the generalized mixed state Renyi entropies (GMSREs), $R_{i,j;i',j'}$ for bosonic field theories.  This development amounts to computing the $n$-point functions that arise in inserting operators at $t=\pm \infty$ on a multi-sheeting Riemann surface - see the Supplemental Material (SM) \cite{SupplementalS1}. Here we combine this development with unitary perturbation theory and truncated spectrum methods to compute $R_2(t)$ at all times after a quench involving two coupled Luttinger liquids.


\noindent\textbf{Quenching from Luttinger Liquids to the Sine-Gordon Model:}
We now want to consider a specific quench, imagining preparing the system in the Luttinger liquid ground state (i.e., taking $J_1$=0 in Eqn. \ref{ei}) and observe the dynamics of the system by turning on at $t=0$ a finite $J_1$.  For $J_1>0$ the dynamics of the system will be that of a far-from equilibrium sine-Gordon model. How far from equilibrium can be quantified.  The energy of the ground state of the sine-Gordon model is 
\begin{eqnarray}
    E_{gs} &=& L\Delta_s^2\tan(\pi\xi/2)/4, ~~~\Delta_s=c(\beta^2)J_1^{(2-\beta^2)^{-1}};\cr\cr
    c(\beta) &=& \frac{2\Gamma(\xi/2)}{\sqrt{\pi}\Gamma(1/2+\xi/2)}\Big(\frac{\pi\Gamma(1-\beta^2/2)}{2\Gamma(\beta^2/2)}\Big)^{1/(2-\beta^2)},
\end{eqnarray}
where $\xi=\beta^2/(2-\beta^2)$ and $\Delta_s$ is the gap of the sine-Gordon soliton excitation.  $c(\beta^2)$ was first determined in \cite{zamo}.
On the other hand the energy of the pre-quench state $|\Phi(t=0)\rangle$ relative to the post-quench Hamiltonian is $-\pi/(6L)$ and so the quench pumps in a finite energy density of $\tan(\pi\xi/2)\Delta_s^2/4$ at large volumes into the system.

The sine-Gordon model is integrable and while integrability does not allow us to determine the non-equilibrium time evolution of the system, it does provide us with knowledge of the dynamically generated non-perturbative scales in the problem.  This include the gap of the solitons, $\Delta_s$, above in terms of $J_1$. It also includes the gaps of solitonic bound states, the breathers.  In sine-Gordon's attractive regime, $\beta<1$, the model has $\left\lfloor\xi^{-1}\right \rfloor$ breathers with gaps
\begin{equation}
    \Delta_{b_n} = 2\Delta_s\sin(\pi n \xi/2), ~~~ n=1,\cdots,\left\lfloor\xi^{-1}\right \rfloor .
\end{equation}
For $\beta \ll 1$, the model has a large number of breathers much lighter than the soliton and it is these excitations that dominate the dynamics.  In this work we will be focusing on the attractive regime and will suppose that $\beta <1$.
With knowledge of these scales, it is possible to write down scaling behavior of various quantities post-quench.  We will focus on both the time-dependent Renyi entropy density as well as the order parameter, $C(t)=\langle \cos(\beta\phi)\rangle(t)$. 

A quantity $O(t)$ with scaling dimension $a$ is going to have a scaling form 
\begin{equation}
    O(t) = \Delta_{b_2}^{a} g_O(\Delta_{b_2} L, \Delta_{b_2} t),
\end{equation}
where $g_0$ is a dimensionless scaling function. For the order parameter $C$, $a=\beta^2$, while for the Renyi entropy densities, $R_n/L$, $a=1$.  We now will determine these scaling forms in the limit of early and late times focusing on the experimentally interesting limit of system sizes $L\Delta_{b_2} \gg 1$. 

\begin{figure}[t]
	\includegraphics[width=0.8\linewidth]{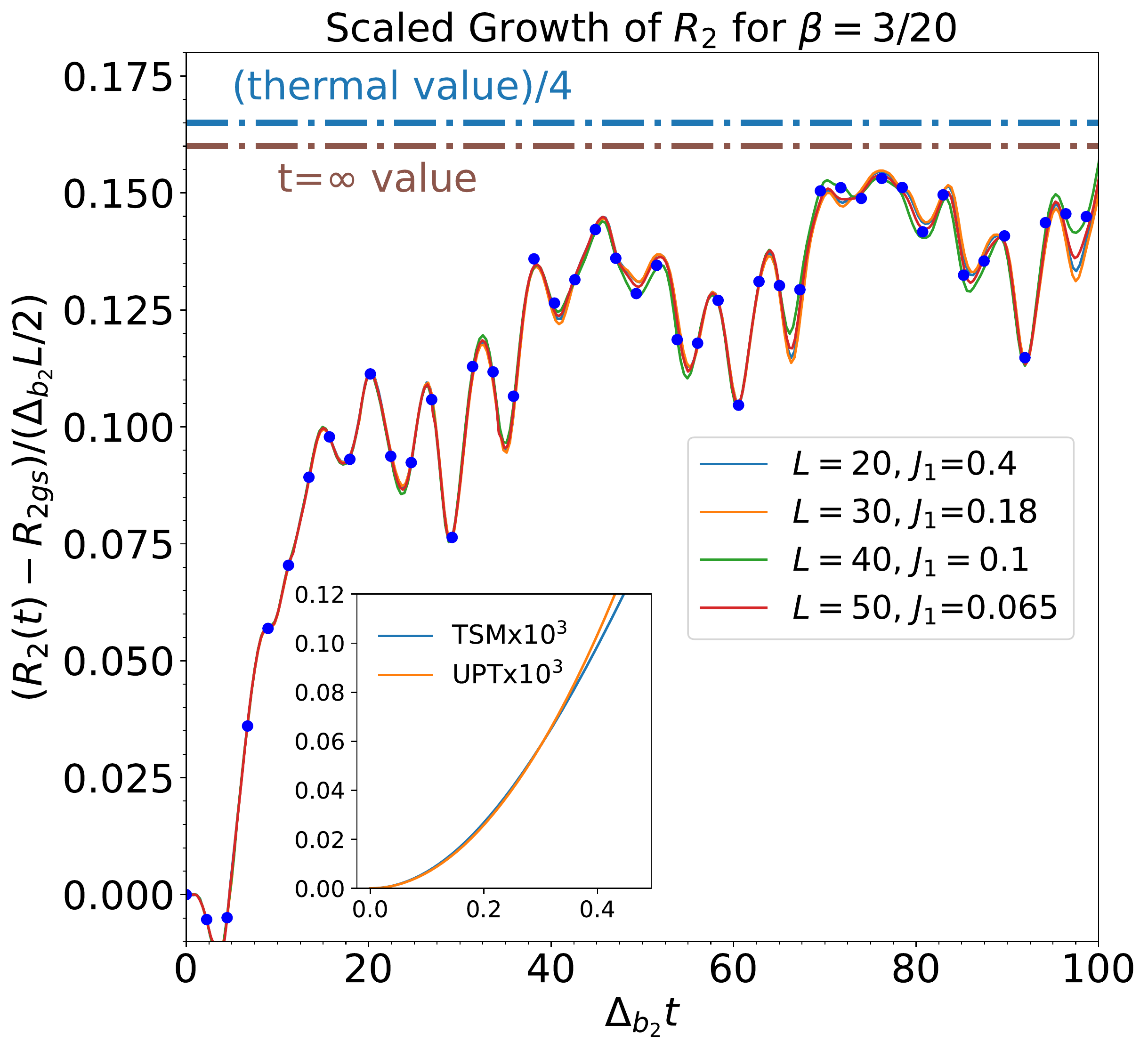}
	\caption{The growth in time of the second Renyi entropy for an equal bi-partition of the system for different system sizes and post-quench couplings $J_1$ chosen such that $\Delta_{b2}L$ is constant and thus scaling collapse is expected. Error bars (blue dots) arising from extrapolation in $E_{c,osc}$ (see S4.4 of the SM) are shown.  Inset: We show the early time behavior of $R_2(t)$ determined by TSM and by UPT - see also S2 of the SM.} \label{scaling_re}
\end{figure}

\noindent\textbf{Early Time Analysis, UPT:}
At early times, we can use unitary perturbation theory (UPT) to determine the leading order term in $J_1$ to the scaling forms.  At the heart of unitary perturbation theory is a similarity transformation that transforms the original unperturbed set of bosonic states to an energy-diagonal one where time evolution is easily evaluated.  In doing so it allows one to derive expressions that are bounded in time \cite{upt}.
Using this framework, the scaling form, $g_O$, simplifies to
\begin{eqnarray}
g_O(x,y) = x^{m(2-\beta^2)-a} h_O(y/x).
\end{eqnarray}
Here $m$ is the order in $J_1$ that gives the leading order correction to $g_0$ in unitary perturbation theory.  For the cosine order parameter, $m=1$, while for the Renyi entropies $m=2$ - see the SM \cite{SupplementalS2}.  While physically less relevant, we also expect this scaling form to hold to arbitrary times in the small volume limit, $L\Delta_{b_2} \ll 1$ as low order UPT becomes increasingly accurate in this limit.  At leading order in UPT, $h_O(z)$, is quadratic in $z$ in all cases.  Thus the initial growth of $C(t)$ and $R_n(t)$ goes as $t^2$.  However at short times $C(t)\sim \beta^2t^2$ while $R_n(t)\sim \beta^4 t^2$.  This difference in the order of $\beta$ reflects how quantum field theoretic the quantity is at short times.  UPT shows that $C(t)$ is determined solely by the zero mode plane wave states $|\nu_i\rangle$ -- that is $C(t)$ at short times is really a quantum mechanical problem of the zero mode, not a field theoretic problem.  The Renyi entropies, $R_n(t)$, in their dependence on a higher order power of $\beta$ directly reflects the presence of the oscillator part of the Bose field.


\noindent\textbf{Longer Time Analysis, TSM:}
While UPT can be used to compute the early time behavior of the growth of $C(t)$ and the Renyi entropies, for longer times we need to use a wholly numerical approach.  The natural choice here is the truncated spectrum methodology (TSM) \cite{yurov1990truncated,yurov1991truncated,James_2018}.  This method provides for a controlled computation of non-equilibrium quantities in a field theoretic setting.  It employs as a computational basis the states of the unperturbed Luttinger liquid, i.e. the $|\Psi_i\rangle$'s, precisely the states for which we now know how to compute the generalized mixed state Renyi entropies.  It gains its name from the need to introduce an energy cutoff, $E_c$, above which we exclude states in the Luttinger liquid basis.  We discuss details of its implementation in the SM \cite{SupplementalS4}.

\begin{figure}[t]
	\includegraphics[width=0.8\linewidth]{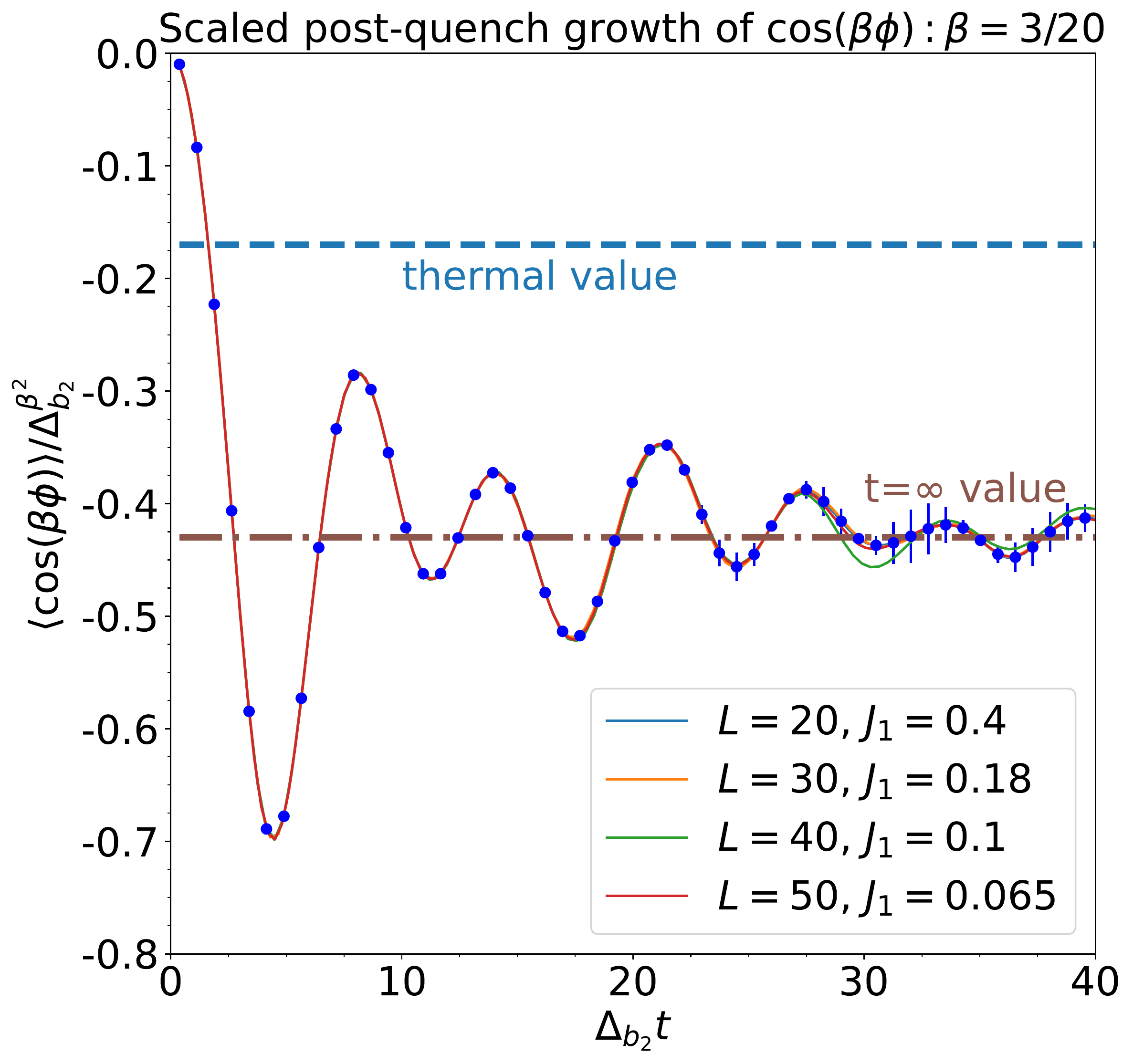}
	\caption{The growth of the order parameter as a function of time for different system sizes and post-quench couplings $J_1$.  We again see the expected scaling collapse.  Error bars (blue lines) arising from extrapolation in $\chi$ are shown.} \label{scaling_op}
\end{figure}

\begin{table*}[t]
\centering
\begin{tabular}{||c|c|c|c|c|c|c|c|c||} 
 \hline
 $\beta$ & $J_1$ & $T_{eff}$ & $(R_2(t=\infty)-R_{2gs})$ & $R_{2,thermal}$ & $\langle \cos(\beta\phi)\rangle(t=\infty)$ & $\langle \cos(\beta\phi)\rangle_{thermal}$ & $R_2$ growth & $\cos(\beta\phi)$ growth \\ 
 \hline \hline
  $3/20$ & 0.1 & 0.45  & $0.16$ & $0.66$ & $-0.43$ & $-0.17$ & $0.009\Delta_{b2}$ & $- 0.003\Delta_{b2}$\\ 
  \hline
  $1/\sqrt{8}$ & 0.0375 & 0.29 & $0.058$ & $0. 19$ & $-0.42 $ & $-0.31$ & $0.005 \Delta_{b2}$ & $-0.023\Delta_{b2}$ \\ 
  \hline
   $1/\sqrt{2}$ & 0.0375 & 0.24 & $0.023$ & $0.093$ & $-0.41$ & $-0.32$ & $0.01 \Delta_{b2}$ & $-0.14 \Delta_{b2}$\\ 
  \hline
\end{tabular}
\caption{Here we report for three values of $\beta$ the late time values of $R_2$ and $\cos(\beta\phi)$, comparing them to their thermal values as determined by the effective temperature $T_{eff}$.  The post-quench values of $J_1$ are chosen such that $\Delta_{b2}(\beta)L$ are constant.  We also report these quantities' early time growth rates.  All values of $R_2,\cos(\beta\phi)$ are scaled by $\Delta_{b2}L/2,\Delta_{b2}^{\beta^2}$.}
\label{growth_and_asympotes}
\end{table*}

As a validation of the accuracy of our TSM results, we demonstrate scaling collapse.  If we fix $L\Delta_{b2}$, we expect data collapse if we plot our post-quench data for $R_2/(\Delta_{b2}L)$ and $\cos(\beta\phi)/\Delta_{b2}^{\beta^2}$ against $t\Delta_{b2}$ for different values of $J_1$ and $L$.  This is what we find, as illustrated in Figs. \ref{scaling_re} and \ref{scaling_op}.  Here we present data that has been extrapolated in the TSM cutoff, $E_c \rightarrow \infty$ (for $R_2$ and $\langle\cos(\beta\phi)$) and the GMSRE exclusion parameter, $W\rightarrow 0$ - see the SM \cite{SupplementalS4}.  If $|\bar\alpha_i\bar\alpha_j\bar\alpha_k\bar\alpha_l|<W$ ($\bar\alpha_i$ is the time-averaged counterpart of $\alpha_i(t)$), we exclude the contribution of $R_{i,j;i',j'}$ to $R_2(t)$ in Eqn. \ref{re_td}.  Because we work with computational bases of size $N_{cb} \sim 10^4$, computing all $R_{i,j;i',j'}$'s would require the computation of $\sim 10^{16}$ different quantities -- something that is computationally prohibitive.  Fortunately the contribution of the vast majority of GMSREs is negligible (because $|\alpha_i(t)\alpha_j(t)\alpha_k(t)\alpha_l(t)|$ is negligible) and we need to only compute a very small fraction of GMSREs in order to compute $R_2(t)$. We conjecture this pattern continues to computing the higher Renyi entropies, $R_n(t)$, i.e. that only a small fraction of the $N_{cb}^{2n}$ GMSREs need to be computed in order to obtain a converged value of $R_n(t)$.  Further details on the extrapolation methods are found in the SM \cite{SupplementalS4}.

\begin{figure}[b]
	\includegraphics[width=0.8\linewidth]{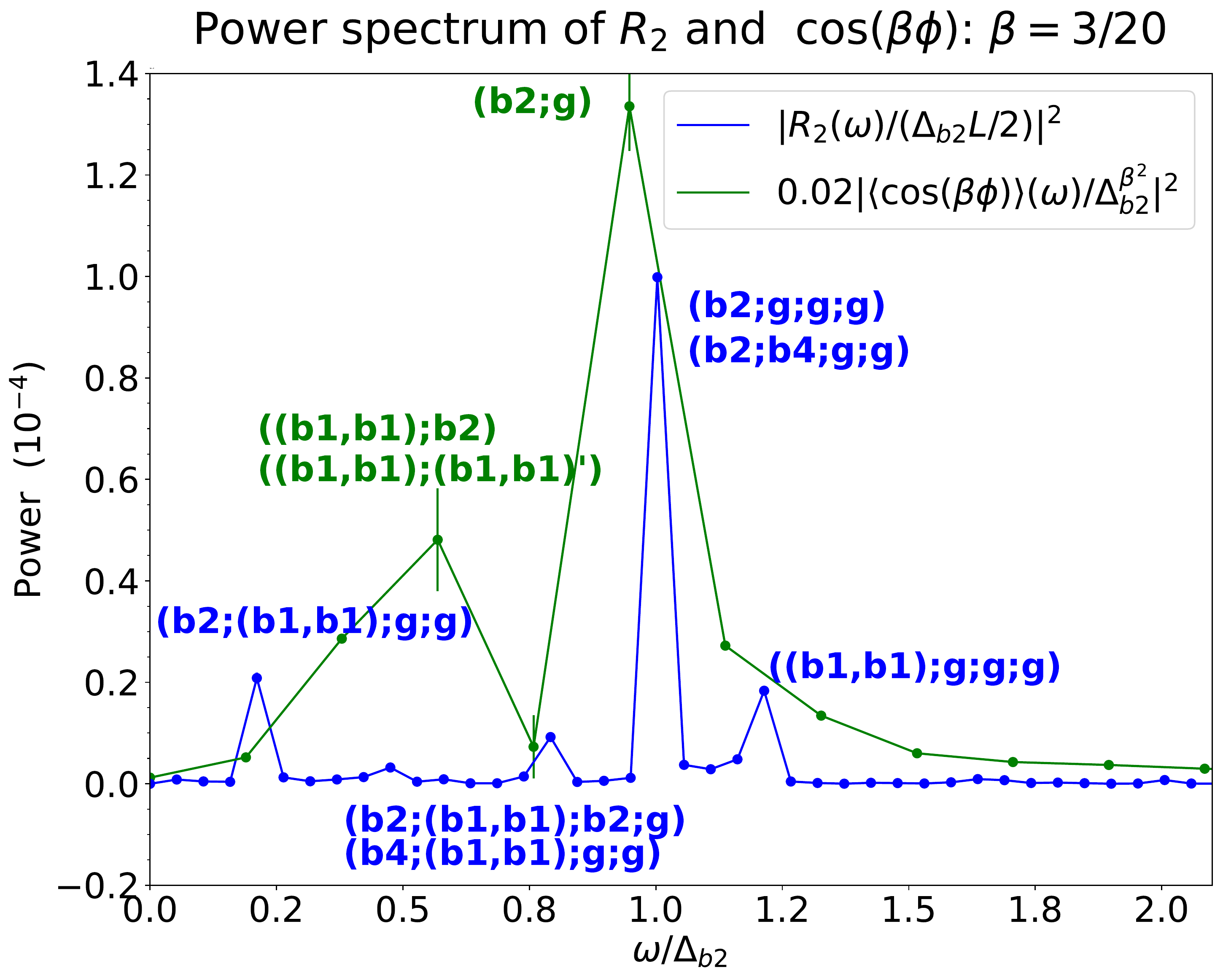}
	\caption{We analyze here for $\beta=0.15$ the oscillation frequencies of the late time behavior of $R_2$ and $\cos(\beta\phi)$ via Fourier transform (FT).  The notation $(e_1;e_2)$ labeling peaks in the FT of $\langle\cos(\beta\phi)\rangle(t)$ indicates a frequency $\omega=|E_{e_1}-E_{e_2}|$ where $E_{e_i}$ is the energy of excitation $e_i$. The $b_i$'s refer to states with single breathers, $(b_i,b_j)$ refers to a state with two breathers, while $g$ is the ground state - see the SM \cite{SupplementalS45}.  Similarly the notation $(e_1,e_2,e_3,e_4)$ appearing in the FT of $R_2(t)$ indicates a frequency $\omega=|E_{e_1}-E_{e_2}+E_{e_3}-E_{e_4}|$.} \label{power}
\end{figure}

The $R_2(t)$ data at $\beta=3/20$ presented in Fig. \ref{scaling_re} shows collapse for four different values of $J_1$ and $L$ (chosen such that $\Delta_{b2}L$ is constant within a few percent) over a time window of $(0,100/\Delta_{b2})$.  We provide error bars associated with the extrapolation procedure.  However for $R_2$ the extrapolation procedure is particularly robust and the error bars are small.  For the collapsed $\langle\cos(\beta\phi)\rangle(t)$ data in Fig. \ref{scaling_op}, we are restricted to a more narrow time window $(0,40/\Delta_{b2})$.  At times $t>40/\Delta_{b2}$, because of dephasing, we cannot reliably extrapolate the order parameter data in $E_c$.  This is reflected in error bars in Fig. \ref{scaling_op} that are visible to the eye for times $t>20/\Delta_{b2}$.

At very early times, UPT predicts quadratic growth in time of $R_2(t)$ and $\langle \cos(\beta\phi)\rangle(t)$.  After UPT breaks down, both of these quantities experience a window in time where they grow linearly.   We report this growth rate in Tab. \ref{growth_and_asympotes} for three different values of $\beta$.  We see that with increasing $\beta$, the growth rates increase in magnitude.

At late times both $R_2(t)$ and $\langle \cos(\beta\phi)\rangle(t)$ saturate. 
We expect $R_2(t)$ to approach its late time value via a correction vanishing as $\log(t)/t^3$, valid for integrable quenches with coherent quasi-particles \cite{Fagotti_Calabrese}.  Using this as a fitting form, we report the value of $R_2(t=\infty)$ in Tab. \ref{growth_and_asympotes}.  We see that asymptotic value of $R_2(t)$ is extremely sensitive to the value of $\beta$.  The late time value of $\langle\cos(\beta\phi)\rangle$ however is not.  We see its final value is almost $\beta$ independent.  Because $\langle\cos(\beta\phi)\rangle (t)$ approaches its asymptote by oscillating about it, its value can be determined most readily by performing a time average over the data obtained after the initial linear growth.  

One useful metric to which we can compare the $t=\infty$ values of $R_2$ and $\cos(\beta\phi)$ are the values that would be obtained if the ensemble governing late time dynamics was thermal.  Because we know the amount of energy injected by the quench, we can use the analytic expression for the energy of the sine-Gordon model arising from the thermodynamic Bethe ansatz (TBA) to compute both the effective temperature that governs the thermal ensemble with this same energy and then the $t=\infty$ values of $R_2$ and $\cos(\beta\phi)$ \cite{rtv-93,km-90,SMTBA}.  We see the expected thermal values of $R_2(t=\infty)$ far exceed that of its post-quench extrapolated value.  Because the sine-Gordon model is integrable, the generalized Gibbs ensemble that governs late time behaviour is going to involve contributions from the higher conserved quantities in the theory.  The system is thus more tightly constrained and so the asymptotic value of the entropy $R_2$ will be smaller than would be expected in a thermal quench.  We also see that the magnitude of $\cos(\beta\phi)$ is in general larger than would be expected from the thermal value.  As this expectation value is directly related to the interaction energy, we can see that the GGE arising from the quench favours interaction over kinetic energy uniformly for different values of $\beta$ in comparison to the thermal ensemble.

As a final comparison between the behavior of $R_2(t)$ and $\cos(\beta\phi)(t)$, we consider the power spectrum of the late time oscillations of these two quantities - see the SM \cite{SupplementalS45}. This is, in effect, a spectroscopic probe of the post-quench Hamiltonian: the frequencies at which power appears here is at the differences of energies of the excitations \cite{delfino-2014,mc-2016} of the post-quench sine-Gordon Hamiltonian.  For $R_2(t)$ these differences involve four excitations while for $\cos(\beta\phi)(t)$ the differences involve two excitations \cite{SM4}.  In Fig. \ref{power} we present the results of the power spectra.  Because of the ability to compute accurately $R_2(t)$ out to longer times, our spectroscopic information for $R_2(t)$ is much resolved in energy than that for $\cos(\beta\phi)(t)$.

\noindent\textbf{Connections to Cold Atomic Systems:} We close this letter by commenting on applications to quenches in cold atomic systems.  The quench considered here (that of joining two Luttinger liquids) has been performed experimentally in Ref. \cite{schmiedmayer2018} while the time evolution of $C(t)$ has been computed in Refs. \cite{dallatorre,horvath}.  Our ability to compute $R_2(t)$ to relatively late times (in comparison to $C(t)$) gives us the time window needed to see equilibration in this system.  At small $\beta$, the equilibration time is 3 to 4 times longer than that needed by $C(t)$ to begin to oscillate about its $t=\infty$ value.  In our spectroscopic analysis of the late time oscillations of $R_2(t)$ and $C(t)$, we can see the outsized role played by the breather excitations of the post-quench Hamiltonian.  Importantly we see the post-quench dynamics cannot be described by the lowest breather alone.  Finally our determination of a $T_{eff}$ for the post-quench dynamics and corresponding thermal values of $R_2(t)$ and $C(t)$ allow us to quantify the importance of the higher conserved quantities in the GGE governing post-quench dynamics.   

\noindent\textbf{Closing Remarks:} In this letter we have presented a general method to compute the time-dependent Renyi entropies, $R_n(t)$, using the notion of a GMSRE, $R_{i,j;i;j'}$, and have applied it to a quantum quench involving the joining of two Luttinger liquids.  Our ability to compute $R_2(t)$ has given us insight into equilibration times in the coupled Luttinger liquid, the importance of the GGE for describing the post-quench dynamics, as well as the importance of the role of higher order breather states that arise because of the non-linear cosine interaction term. 
We mention that while the quasi-particle picture \cite{cc-2005} provides the exact time evolution of the von Neumann entropy ($n=1$) for arbitrary integrable models \cite{ac-2017,ac-2018}, the same is not true \cite{ac2-2017,ac3-2017,acm-2018,klobas2021entanglement,bertini2022} for the experimentally accessible Renyi entropies for which our approach is the only viable methodology for both integrable and chaotic post-quench dynamics.

\noindent Note added: After the submission of this manuscript a related work appeared \cite{https://doi.org/10.48550/arxiv.2202.11113} in which a different method to compute entanglement in bosonic quantum field theories appeared.
\acknowledgements
 R.M.K. was supported by the U.S. Department of Energy, Office of Basic Energy Sciences, under Contract No. DE-AC02-98CH10886. 
 P.C. and S.M. acknowledge support from the ERC under Consolidator grant number 771536 (NEMO).  

\bibliography{bib}
\clearpage
\newpage

\setcounter{equation}{0}
\setcounter{figure}{0}
\setcounter{table}{0}
\makeatletter
%
%
%
%


\beginsupplement


\noindent Here we report  some additional technical details on our work. In particular, in Section \ref{SI:Generalized Renyi Entropies} we report the form of the generalized mixed state Renyi entropies. Then we compute the postquench early time behaviour of $R_2(t)$ and $C(t)$ using the unitary perturbation theory in Section \ref{SI:Unitary Perturbation Theory}. In Section \ref{SI:TBA}, we give some details of the thermodynamic Bethe ansatz for the sine-Gordon model. Finally, we describe the truncated spectrum methods, the extrapolation procedures, the power spectrum and some additional data with respect to the ones in the main text in the last section \ref{SI:TSM}.
\section{Generalized Mixed State Renyi Entropies}
\label{SI:Generalized Renyi Entropies}
\begin{figure}[b]
	\includegraphics[width=0.6\linewidth]{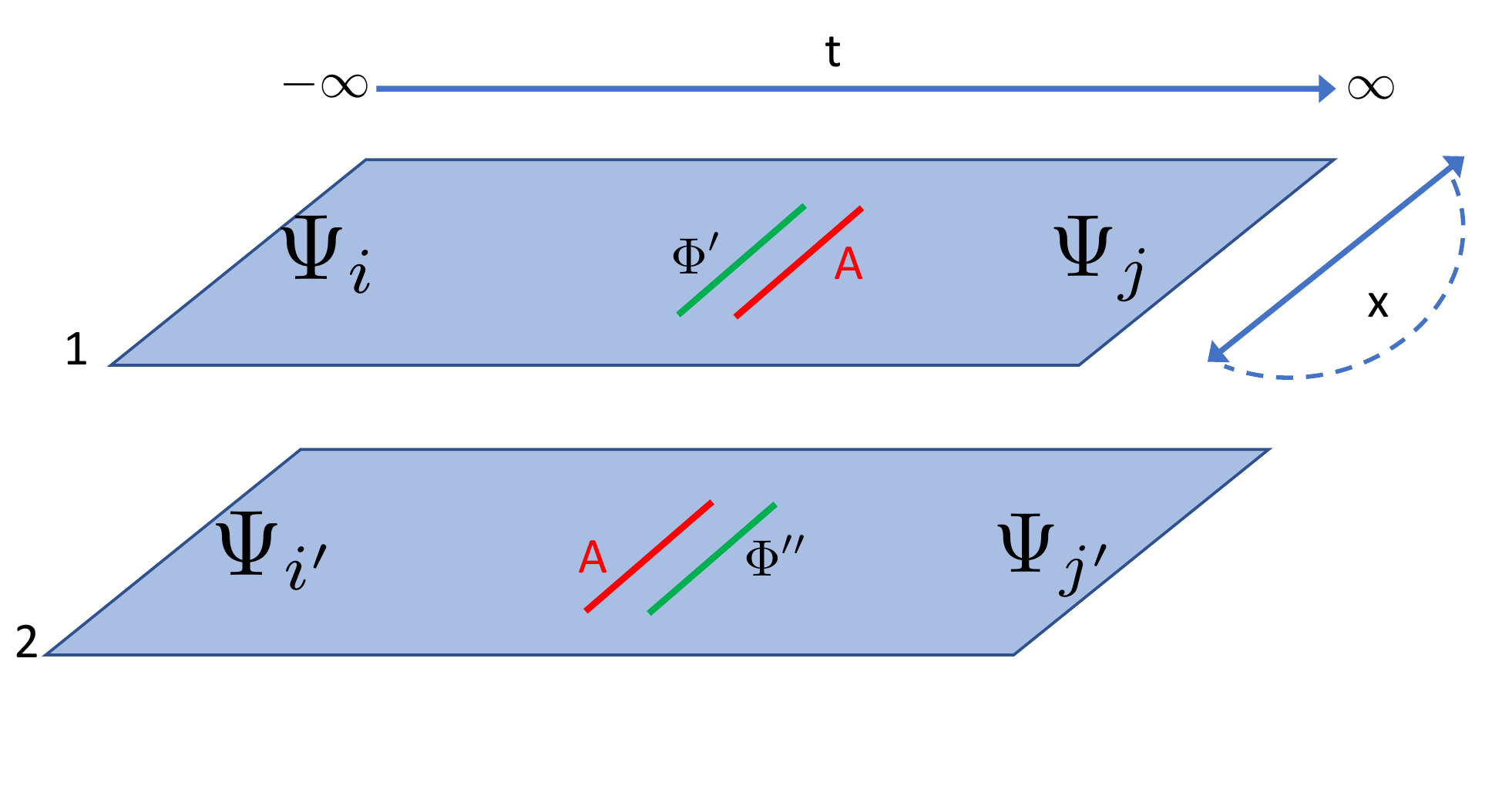}
	\caption{The space-time by which a generalized second Renyi entropy is computed.  The green and red lines in the two-sheeted Riemann surface are identified.  The red line corresponds to the part of the system, A, left after tracing out region B.} \label{spacetime_R2}
\end{figure}

In this section we record the form of the generalized mixed state Renyi entropies (GMSREs) for arbitrary bosonic states.  This form is derived in Ref. \onlinecite{long_paper}.  The GMSRE for the second Renyi entropy is defined in terms of a 4-tuplet of states,
\begin{equation}
R_{1,2;3,4} = {\rm Tr}_A ({\rm Tr}_B |\Psi_{1}\rangle\langle \Psi_{2}|{\rm Tr}_B |\Psi_{3}\rangle\langle \Psi_{4}| ,
\end{equation}
where each state $|\Psi_{1,2,3,4}\rangle$ is defined by its oscillator and vertex operator content and can be factorized into a left and right moving piece:
\begin{eqnarray}
    |\Psi_i\rangle &=& |\Psi_{Li}\rangle|\Psi_{Ri}\rangle; \cr\cr  |\Psi_{Li}\rangle &=& \prod^{N_i}_{j=1}a_{-k^{(i)}_j}| \nu_i\rangle;\cr\cr
    |\Psi_{Ri}\rangle &=& \prod^{\bar N_i}_{j=1}\bar a_{-\bar k^{(i)}_j} |\bar \nu_i\rangle;\cr\cr
    |\nu_i\rangle \equiv e^{i\nu_i\beta \phi_L}|0\rangle, && ~~~|\bar\nu_i\rangle \equiv e^{i\nu_i \phi_R}|0\rangle,
\end{eqnarray}
where $\nu_i$ is an integer-multiple of $\beta$.
Evaluating $R_{1,2;3,4}$ amounts to computing a four point function on a spacetime consisting of a two-sheeted Riemann surface (see Fig.\ref{spacetime_R2}).  Thus like with any conformal correlator, $R_{1,2;3,4}$ can be written as a product of a chiral piece and an anti-chiral piece:
\begin{eqnarray}
R_{1,2;3,4} &\equiv& R^L_{1,2,3,4}R^R_{1,2,3,4};\cr\cr
R^L_{1,2,3,4} &=& \langle \Psi_{Li}(t=-\infty)\Psi_{Lj}(t=\infty)\Psi_{Li'}(t=-\infty)\Psi_{Lj'}(t=\infty)\rangle
\equiv R^{\nu_1,\nu_2,\nu_3,\nu_4}_{k_1,\dots,k_N};\cr\cr
R^R_{1,2,3,4} &=& \langle \Psi_{Ri}(t=-\infty)\Psi_{Rj}(t=\infty)\Psi_{Ri'}(t=-\infty)\Psi_{Rj'}(t=\infty)\rangle\equiv R^{\nu_1,\nu_2,\nu_3,\nu_4}_{\bar k_1,\dots,\bar k_{\bar N}} ,
\end{eqnarray}
where the tuplets $(k_1,\cdots,k_N)$ and $(\bar k_1,\cdots,\bar k_N)$ are defined by
\begin{eqnarray}
(k_1,\cdots,k_N) &=& (k^{(1)}_1,\cdots,k^{(1)}_{N_1},k^{(2)}_1,\cdots,k^{(2)}_{N_2},k^{(3)}_1,\cdots,k^{(3)}_{N_3},k^{(4)}_1,\cdots,k^{(4)}_{N_4})\cr\cr
(\bar k_1,\cdots,\bar k_{\bar N}) &=& (\bar k^{(1)}_1,\cdots,\bar k^{(1)}_{\bar N_1},\bar k^{(2)}_1,\cdots,\bar k^{(2)}_{\bar N_2},\bar k^{(3)}_1,\cdots,\bar k^{(3)}_{\bar N_3},\bar k^{(4)}_1,\cdots,\bar k^{(4)}_{\bar N_4}),
\end{eqnarray}
and $N=\sum^4_{i=1}N_i, \bar N=\sum^4_{i=1}\bar N_i$

Because the right and left parts of $R_{1,2;3,4}$ can be identified up to a complex conjugation, we focus on the left moving piece,  $R^{\nu_1,\nu_2,\nu_3,\nu_4}_{k_1,\dots,k_N}$.  This quantity is given by\cite{long_paper}:

\begin{multline}\label{eq:vv}
\frac{R^{\nu_1,\nu_2,\nu_3,\nu_4}_{k_1,\dots,k_N}}{R_{\mathbbm{1},\mathbbm{1},\mathbbm{1},\mathbbm{1}}}=M(\nu_1,\nu_2,\nu_3,\nu_4)A_1A_2A_3A_4(-1)^{N_1+N_3}e^{2\pi i\frac{v}{R}(P_1+P_3-P_2-P_4)}\Big[ F_{k_1,\dots,k_N}\\
+\sum_{i}^4 F_{k_1,\dots, \widehat{k_{i}} \dots,  k_N} L_{k_i}(\bar{\nu}) +\sum_{i_1<i_2}^4 F_{k_1,\dots, \widehat{k_{i_1}} \dots, \widehat{k_{i_2}} \dots k_N} L_{k_{i_1}}(\bar{\nu})L_{k_{i_2}}(\bar{\nu}) \\
+\sum_{i_1<i_2<i_3}^4F_{k_1,\dots, \widehat{k_{i_1}} \dots, \widehat{k_{i_2}}, \dots\widehat{k_{i_3}} \dots  k_N} L_{k_{i_1}}(\bar{\nu})L_{k_{i_2}}(\bar{\nu})L_{k_{i_3}}(\bar{\nu})   +\dots + 
 \prod_{i=1}^N L_{k_{i}}(\bar{\nu}) 
\Big ].
\end{multline}
Let us try to understand better each component of this involved equation. $R_{\mathbbm{1},\mathbbm{1},\mathbbm{1},\mathbbm{1}}$ is the second R\'enyi entropy of the ground state of the system. $M(\nu_1,\nu_2,\nu_3,\nu_4)$ encodes information about the vertex operator part of the generalized R\'enyi entropies and it can be computed in terms of $\nu_i$'s and the ratio between the subsystem size, $\ell$, and the system size, $L$, $r=\ell/L$ :
\begin{multline}
M(\nu_1,\nu_2,\nu_3,\nu_4)
=\sin \left(\frac{\pi r}{2} \right)^{\nu_1\nu_2+\nu_3\nu_4}\cos \left(\frac{\pi r}{2} \right)^{\nu_1\nu_4+\nu_2\nu_3} \\ \times2^{-\bar{M}\cdot \bar{M}/2}\sin(\pi r)^{\bar{M}\cdot \bar{M}/2}  (e^{-i\pi r}e^{2\pi i v/R})^{(\nu_1^2+\nu_3^2-\nu_2^2-\nu_4^2)/2},
\end{multline}
where $M=(\nu_1,\nu_2,\nu_3,\nu_4)$.
$A_j$'s denote the normalization of each of the four states 
\begin{equation}
A_j=1/(\langle 0|\prod_{i=1}^{N_j}a_{k^{(j)}_i}\prod_{i=1}^{N_j}a_{-k^{(j)}_i}|0\rangle, ~~j=1,2,3,4.
\end{equation}
The terms $F_{k_1,\dots,k_N}$ are given in terms of Hafnians and read
\begin{equation}\label{eq:n2}
F_{k_1,\dots,k_N}=\sum_{\substack{\sigma \in S_N\\ \sigma_{2i}<\sigma_{2i+1}\\ \sigma_{1}< \sigma_{3}\cdots \sigma_{2N-1}}}\prod_{i=1}^{N/2}W(k_{\sigma_{2i-1}},k_{\sigma_{2i}},y_{\sigma_{2i-1}},y_{\sigma_{2i}}),
\end{equation}
where $S_N$ is the permutation group and
\begin{equation}\label{eq:Wk's}
\begin{split}
W(k_i, k_j , y_i, y_j )&=\begin{cases}
\frac{1}{\Gamma(k_i)}\sum_{l=0}^{k_i-1} {{k_i-1}\choose{l}} \frac{\Gamma(k_i-l+1)}{\Gamma(k_i+k_j-l+1)}  (\partial_z^l f^{k_i})(z=y_i,y_i)\\ \qquad \times (\partial_z^{k_i+k_j-l}f^{k_j})(z=y_j,z=y_j), ~~~ y_i=y_j; \\
\frac{1}{\Gamma(k_i)\Gamma(k_j)}\partial_{z_i}^{k_i-1}\partial_{z_j}^{k_j-1}\left(\frac{f^{k_i}(z_i,y_i)f^{k_j}(z_j,y_j)}{(z_i-z_j)^2} \right)\Big |_{\substack{z_i=y_i \\ z_j=y_j}}, ~~~~ y_i\neq y_j;
\end{cases}\\
f(z_i,y_j)&=\frac{z_i^2-(y_j^*)^2}{z_i+y_j}.
\end{split}
\end{equation}
The notation $F_{k_1,\dots, \widehat{k_{i'}}, \dots k_N}$ indicates that the sequence of modes $k_1\dots k_N$ does not contain $k_{i'}$.
The set of points $y_i$, $i = 1,\dots , N$, are defined as
\begin{equation}
y_i=\begin{cases}
e^{i\pi r/2}, & 1\leq i \leq N_1; \\
e^{-i\pi r/2}, & N_1+1 \leq i \leq N_1+N_2; \\
-e^{i\pi r/2}, & N_1+N_2+1 \leq i \leq N_1+N_2+N_3;\\
-e^{-i\pi r/2}, & N_1+N_2+N_3 +1\leq i \leq N_1+N_2+N_3+N_4,
\end{cases}
\end{equation}
while $P_i$, the total (chiral) momentum of $|\Psi_i\rangle$ is given by
\begin{equation}
P_i=\sum_{j=1}^{N_i}k^{(i)}_{j},  \quad i=1\dots 4.
\end{equation}

Finally, the terms $L_{k_j}(\bar{\nu})$, $j=1, \cdots N$ appearing in Eqn.\ref{eq:vv} can be written in terms of the function $f(z_i,y_j)$ as
\begin{equation}
\begin{split}
    L_{k_j}(\bar{\nu})=&\sum_{i=1}^4 \nu_i J_{ij},\\
    J_{ij}=& \begin{cases}
   \frac{1}{\Gamma(k_j+1)}\partial_{z_j}^{k_j}f^{k_j}(z_j,y_j)  & i=j;\\
   \frac{1}{\Gamma(k_j)}\partial_{z_j}^{k_j-1}\frac{f^{k_j}(z_j,y_j) }{z_j-y_i} & i\neq j.
    \end{cases}
    \end{split}
\end{equation}


\section{Unitary Perturbation Theory}
\label{SI:Unitary Perturbation Theory}

In this section, we want to compute the time evolution of observables after a quantum quench that do not commute with the $t < 0$ Hamiltonian.  We do so by adapting the unitary perturbation theory (UPT)of Ref. \onlinecite{upt}. This formalism allows us to analytically compute the postquench early time behaviour of the second and third R\'enyi entropies as well as the order parameter $C(t)=\langle \cos(\beta\phi)\rangle(t)$. 

The main idea of this formalism is to bring the Hamiltonian into energy diagonal form.  To do so we introduce a canonical anti-Hermitian transformation,
\begin{equation}\label{eq:a-h}
S=J_1 S_1 +\frac{J_1^2}{2}S_2 +O(J_1^3).
\end{equation}
We will apply it to the Hamiltonian
\begin{equation}
\begin{split}
H&=H_0+H_1, \\
H_0&=\frac{2\pi}{L}\left[ \sum_k(a_{-k}a_k+\bar{a}_{-k}\bar{a}_k)+\pi_0^2-\frac{1}{12}\right], \\
H_1&=J_1 \left(\frac{2\pi}{L}\right)^{\beta^2}\int_0^L dx :\cos(\beta \phi(x)):,
\end{split}
\end{equation}
where $:\dots:$ denotes the standard normal ordering prescription.\cite{CFTBook} 

The action of $S$ upon $H$ in Eqn. \eqref{eq:a-h} is given by
\begin{equation}
\begin{split}
e^{-S}He^{-S}=&H_0+J_1(H_1+[S_1,H_0])+J_1^2
(\frac{1}{2}[S_2,H_0]+[S_1,H_1]+\frac{1}{2}[S_1,[S_1,H_0]])+O(J_1^3),\\
\equiv & H_0+J_1 H_{1,diag}+J_1^2H_{2,diag}+O(J_1^3).
\end{split}
\end{equation}
We define $S$ such that the matrix elements of $H_{n,diag}$ with respect to two eigenstates, $|n\rangle, |m\rangle$, of $H_0$ are only non-zero if $E_n=E_m$. Hence, we find that $S$ satisfies at first order in $J_1$,
\begin{equation}
S_{1,nm}=\begin{cases}
\frac{H_{1,nm}}{E_n-E_m} \quad &E_n\neq E_m,\\
0 \quad  &E_n= E_m,
\end{cases}
\end{equation}
and at second order, 
\begin{equation}
S_{2,nm}=\begin{cases}
\frac{[S_1,H_1+H_{1,diag}]}{E_n-E_m} \quad &E_n\neq E_m,\\
0 \quad  &E_n= E_m.
\end{cases}
\end{equation}
The transformed Hamiltonian $H_{1/2,diag}$ reads to second order in $J_1$,
\begin{equation}
\begin{split}
\langle n|H_{1,diag}|m\rangle =& \langle n|H_{1}| m \rangle; ~~~ E_n=E_m \\
\langle n|H_{2,diag}|m\rangle =& \sum_{k,E_k\neq E_n} \frac{H_{1,nk}H_{1,km}}{E_n-E_k}, ~~~E_n=E_m.
\end{split}
\end{equation}
We can apply this formalism to find the time dependence of an observable $A$
\begin{equation}
\begin{split}
\braket{A(t)}=&\braket{0|e^{iHt}Ae^{-iHt}|0}\\
=& \braket{0|e^{-S}e^{iH_{diag}t}e^SAe^{-S}e^{-iH_{diag}t}e^S|0}\\
\equiv & \braket{0|e^{-S}e^{S(t)}A_{diag}(t)e^{-S(t)}ee^S|0},
\end{split}
\end{equation}
where $S(t)=e^{iH_{diag}t}Se^{-iH_{diag}t}$, $A_{diag}(t)=e^{iH_{diag}t}Ae^{-iH_{diag}t}$. We expand first the inner transformation as
\begin{equation}
e^{S(t)}A_{diag}(t)e^{-S(t)}=A_{diag}(t)+[S(t),A_{diag}(t)]+\frac{1}{2}[S(t),[S(t),A_{diag}(t)]]+O(J_1^3),
\end{equation}
and then the outer back transformation
\begin{multline}
e^{-S}e^{S(t)}A_{diag}(t)e^{-S(t)}e^S=A_{diag}(t)+[S(t)-S,A_{diag}(t)]\\+\frac{1}{2}\left([S,[S-2S(t),A_{diag}(t)]]+[S(t),[S(t),A_{diag}(t)]]\right)+O(J_1^3).
\end{multline}
Therefore we can write down $\braket{A(t)}$ in terms of its matrix elements as
\begin{multline}\label{eq:upt1}
\braket{A(t)}=A_{00}+\sum_j\left(\delta S_{0k}(t) A_{diag,k0}(t)-A_{diag,00}(t)\delta S_{k0}(t)\right) \\
\hskip -.3in +\frac{1}{2}\sum_{kl}\left(\delta S_{0k}(t)\delta S_{kl}(t)A_{diag,l0}(t) +A_{diag,0k}(t) \delta S_{kl}(t)\delta S_{l0}(t) \right.\\
\left. \hskip -.3in - 2\delta S_{0k}(t)A_{diag,kl}(t)\delta S_{l0}(t)\right) +O(J_1^3), 
\end{multline}
with $\delta S_{kl}(t)\equiv S_{kl}(t)-S_{kl}$. \\ 

We now use UPT to compute the time-dependence of the state of the system, $|\Phi(t)\rangle$.   To do so, we write $|\Phi(t)\rangle$ in terms of the states of the unperturbed bosonic theory $|\Psi_a\rangle$ via:
\begin{equation}\label{eiv}
|\Phi(t)\rangle = \sum_a \alpha_a(t)|\Psi_a\rangle.
\end{equation}
By choosing the observable $A$ as $\rho_{ab}=\ket{\Psi_a}\bra{\Psi_b}$, we can use Eq. \eqref{eq:upt1} to compute the time dependence of the density matrix elements 
\begin{equation}
c_{ab}(t) \equiv \langle \Phi(t)| \rho_{ab}|\Phi(t)\rangle
\end{equation}
to second order in $J_1$ as
\begin{equation}\label{eq:cab}
c_{ab}(t)=\begin{cases}
1+2\sum_{k} S_{1,0k}S_{1,k0}(1-\cos((E_k-E_0) t)) & a=b=0 \\
 e^{it(E_0-E_b)} S_{1,b0} -  S_{1,b0}\\ \hskip .5in +\frac{1}{2}\sum_kS_{1,bk}S_{1,k0}(1-e^{i(E_0-E_k)t}-e^{i(E_k-E_b)t}+e^{i(E_0-E_b)t}) & a=0, b \neq 0 \\
  -e^{it(E_a-E_0)} S_{1,0a} +  S_{1,0a} +\\
  \hskip .5in \frac{1}{2} \sum_kS_{1,0k}S_{1,ka}(1-e^{i(E_k-E_0)t}-e^{i(E_a-E_k)t}+e^{i(E_a-E_0)t}) & a\neq 0, b = 0 \\
 - (1-e^{i(E_0-E_b)t}-e^{i(E_a-E_0)t}+e^{i(E_a-E_b)t})S_{1,0a}S_{1,b0} & a,b \neq 0 \\
\end{cases}
\end{equation}
This will allow us to back out the $\alpha_a(t)$'s.

At small $\beta$, the number of states we need to consider in the post-quench density matrix at leading order in the cosine coupling, $J_1$, and leading order in $\beta$ include 
\begin{equation}
\begin{split}\label{eq:statesn2}
&\ket{0;0;0}\equiv |0\rangle,~~ \ket{0;0;m=\pm1}\equiv e^{im\beta \phi (0)}\ket{0},~~ \ket{n;n;m=\pm 1}\equiv \frac{1}{n}a_{-n}\bar{a}_{-n}e^{im\beta \phi (0)}\ket{0},\\
&\ket{n,l;n,l;m=0,\pm 1}\equiv \frac{1}{2^{\delta_{nl}}n l}a_{-n}a_{-l}\bar{a}_{-n}\bar{a}_{-l}e^{im\beta \phi (0)}\ket{0},\\
&\ket{n,l;n+l;m=\pm 1}\equiv \frac{1}{2^{\delta_{nl}/2}\sqrt{n l (n+l)}}a_{-n}a_{-l}\bar{a}_{-n-l}e^{im\beta \phi (0)}\ket{0},\\
&\ket{n+l;n,l;m=\pm 1}\equiv \frac{1}{2^{\delta_{nl}/2}\sqrt{n l (n+l)}}a_{-n-l}\bar{a}_{-n}\bar{a}_{-l}e^{im\beta \phi (0)}\ket{0}.
\end{split}
\end{equation}
The energies of these states are given by 
 \begin{equation}
 E_{(n_1,\cdots,n_r);(l_1,\cdots,l_s);m}=\frac{2\pi}{L}(\sum^r_{i=1}n_i+\sum^s_{i=1}l_i+m^2\beta^2).
 \end{equation}
We will further focus on the contribution of states involving chiral modes such that $\sum_{i=1}^2 n_i \leq 2$ as these states provide the dominant contribution to $R_2(t)$.
Thus we consider the contribution of states: $\ket{n;n;0,\pm 1}, n=0,1,2; \ket{1,1;2;\pm 1};\ket{2;1,1;\pm 1}$, and $\ket{1,1;1,1;0,\pm 1}$. Using Eq. \eqref{eq:cab}, the coefficients, $\alpha_{n,m}(t)$, describing these states' time dependence post-quench are:
\begin{equation}\label{eq:coeff}
\begin{split}
\alpha_{n;n;\pm 1}(t)&=\left( \frac{e^{it(E_{n;n;1}-E_{0;0;0})}-1}{E_{n;n;1}-E_{0;0;0}} \right)J_1\frac{\beta^2}{n} L\left(\frac{2\pi}{L} \right)^{\beta^2}\simeq iJ_1\frac{\beta^2}{n} L\left(\frac{2\pi}{L} \right)^{\beta^2} t;\\
\alpha_{0;0;\pm 1}(t)&=\left( \frac{e^{it(E_{0;0;1}-E_{0;0;0})}-1}{E_{0;0;1}-E_{0;0;0}} \right)J_1 L\left(\frac{2\pi}{L} \right)^{\beta^2} \simeq iJ_1 L\left(\frac{2\pi}{L} \right)^{\beta^2} t;\\
\alpha_{0;0;0}(t)&=1-2J_1^2 L^2\left(\frac{2\pi}{L} \right)^{2\beta^2}\left[\frac{1-\cos(E_{0;0;\pm1} t)}{ E^2_{0;0;\pm1}} +\sum_{k=1}^2 \frac{\beta^4}{k^2}\frac{1-\cos(E_{k;k;\pm1} t)}{ E^2_{k;k;\pm1}}\right]\\
&\simeq 1-J_1^2 L^2t^2\left(\frac{2\pi}{L} \right)^{2\beta^2}\left[ 1+\sum_{k\neq 0} \frac{\beta^4}{k^2}\right]; \\
a_{1,1;1,1;\pm 1}(t)&=\frac{1}{2}\left( \frac{e^{it(E_{1,1;1,1;1}-E_{0;0;0})}-1}{E_{1,1;1,1;1}-E_{0;0}} \right)J_1\beta^4 L\left(\frac{2\pi}{L} \right)^{\beta^2}\simeq\frac{1}{2} i J_1 \beta^4 L\left(\frac{2\pi}{L} \right)^{\beta^2}t; \\
\alpha_{1,1;1,1;0}(t)&=\frac{1}{4}\sum_{E_k\neq 0, E_{1,1;1,1;0}}\Big[ \frac{H_{1,(1,1;1,1;0)k}H_{k0}}{(E_k-E_{0;0})(E_{1\,1;1,1;0}-E_k)}(1-e^{-i(E_k-E_{0;0})t}-e^{i(E_k-E_{1\,1;1,1;0})t}\\ &+e^{-i (E_{1\,1;1,1;0}-E_{0;0})t})\Big]
\simeq -3J_1^2 L^2t^2\left(\frac{2\pi}{L} \right)^{2\beta^2}\beta^4; \\
\alpha_{1,\, 1;2; 1}(t)&=-a_{2;1,\, 1;-1}(t) \simeq\frac{1}{2\sqrt{2}} i J_1 \beta^3 R\left(\frac{2\pi}{R} \right)^{\beta^2}t.
\end{split}
\end{equation}
With these coefficients in hand, we can plug them into our generalized Renyi entropy machinery to compute the time-dependence of the second R\'enyi entropy.

\begin{figure}[b]
	\includegraphics[width=0.6\linewidth]{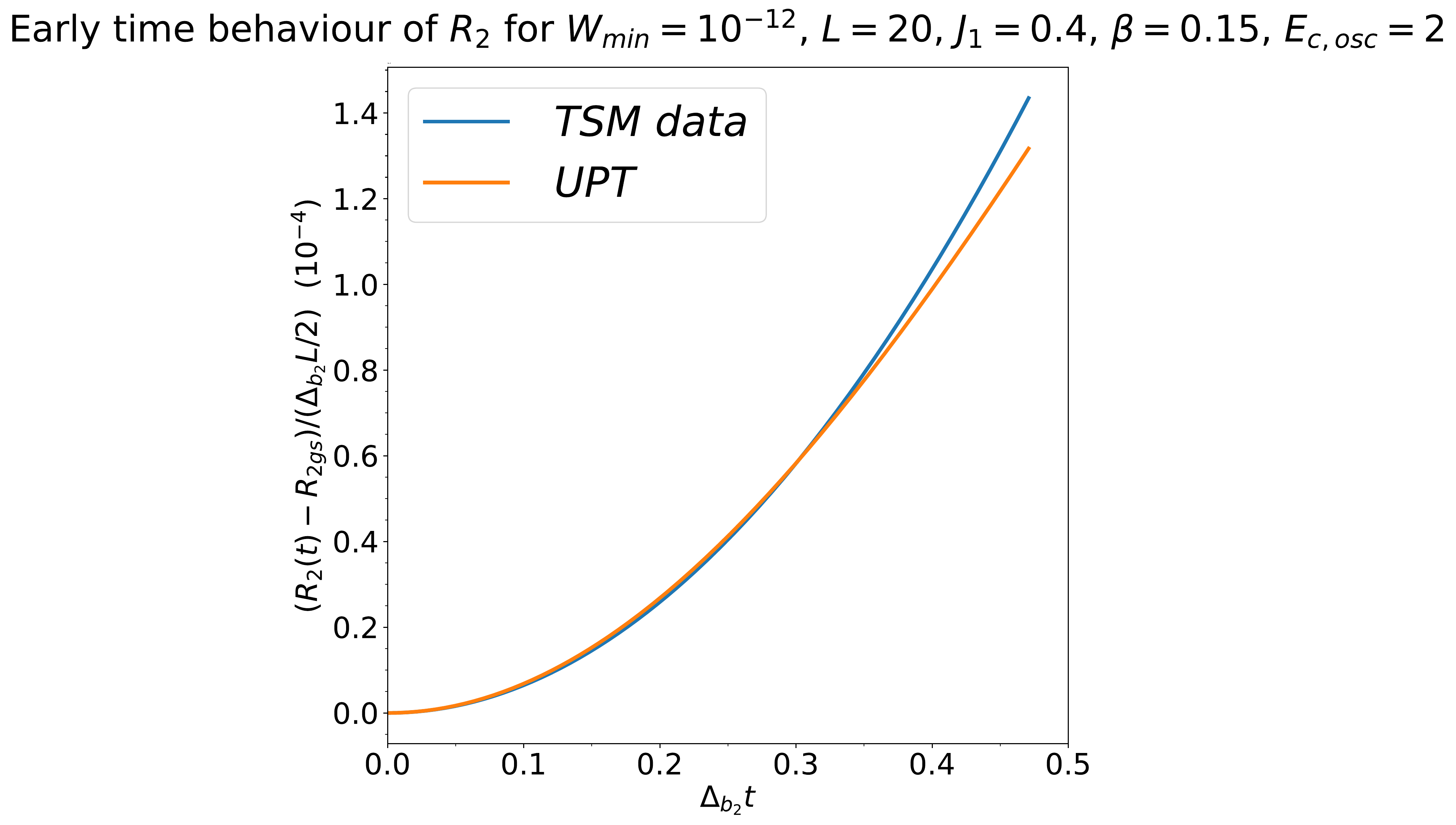}
	\caption{Early-time growth of $R_2(t)-R_{2,gs}$ computed through Truncated Spectrum Methods (blue) and Eq. \eqref{eq:n2} (orange). The TSM data has been obtained by choosing the cutoff $E_{c,osc}=2$ (see Sec. \ref{subsec:cutoff}). We see at early times a good match between the TSM data and the UPT.} \label{early_time}
\end{figure} 

In order to perform this computation analytically, we need the following non vanishing (non-chiral) generalized Renyi entropy and all their possible permutations.  These are given in Table \ref{gen_ren}. Putting everything together, we find
\begin{eqnarray}\label{eq:n2}
R_2(t)- R_{2,gs} &=&
L^2 J^2_1 \left(\frac{2\pi}{L} \right)^{2\beta^2}t^2\beta^4\Big(4\log^2(2) -\frac{447}{256}\Big) + O(t^3)\cr\cr
&=& 0.17 L^2 J^2_1 \left(\frac{2\pi}{L} \right)^{2\beta^2}t^2\beta^4 + O(t^3).
\end{eqnarray}
We have kept only terms up to $O(\beta^4)$, and we neglected the contributions due to states $\ket{1,1;2;\pm 1}$ because they are $O(\beta^6)$. We see that $R_2(t)$ behaves as $\beta^4 t^2$.  The $\beta^4$ dependence of $R_2(t)$ means that we cannot ignore the contribution of states of the form $\ket{n>0;l>0;m}$, i.e. states with a non-trivial bosonic mode, $a_{-n},\bar a_{-n}$, content. Thus $R_2(t)$ probes at early times not just the zero mode dynamics of the field, $\phi(t)$, but its field theoretic nature.

In Fig. \ref{early_time} we compare this analytical prediction with the TSM data using a cutoff $E_{c,osc}=2$ (see Sec. \ref{subsec:cutoff} for more details about this parameter). This TSM data includes contributions of generalized Renyi entropies involving quadruplets with more than two chiral modes and $m>1$ in Eq. \eqref{eq:statesn2}. Therefore the two curves do not overlap, but Eq. \eqref{eq:n2} provides a good approximation of the data at early times.

\begin{table*}[t]\label{gen_ren}
\centering
\begin{tabular}{||c|c|c|c|c|c||} 
 \hline
$\ket{1}$ & $\ket{2}$ & $\ket{3}$ & $\ket{4}$ &  $R_{1,2;3,4}/R_{\mathbbm{1},\mathbbm{1},\mathbbm{1},\mathbbm{1}}$ & multiplicity  \\ 
 \hline \hline
  $|0;0;1\rangle$ & $|0;0;-1\rangle$ & $|0;0;0\rangle$ & $|0;0;0\rangle$ & $2^{-\beta^2}$ & 8 \\ 
   \hline
   $|0;0;1\rangle$ & $|0;0;0\rangle$ & $|0;0;-1\rangle$ & $|0;0;0\rangle$ & $2^{-2\beta^2}$ & 4 \\ 
   \hline
   $|1;1;1\rangle$ & $|0;0;-1\rangle$ & $|0;0;0\rangle$ & $|0;0;0\rangle$ & $2^{-\beta^2}\left(\frac{\beta^2}{2}\right)^{2}$ & 8 \\ 
   \hline
   $|2;2;1\rangle$ & $|0;0;-1\rangle$ & $|0;0;0\rangle$ & $|0;0;0\rangle$ & $2^{-\beta^2}\left(\frac{\beta^2}{4\sqrt{2}}\right)^{2}$ & 8 \\ 
   \hline
   $|2;2;1\rangle$ & $|0;0;0\rangle$ & $|0;0;-1\rangle$ & $|0;0;0\rangle$ & $2^{-2\beta^2}\left(\frac{\beta^2}{2\sqrt{2}}\right)^{2}$ & 4 \\ 
   \hline
   $|1;1;1\rangle$ & $|1;1;-1\rangle$ & $|0;0;0\rangle$ & $|0;0;0\rangle$ & $2^{-\beta^2}\left(1/2+\frac{\beta^2}{4}\right)^{2}$ & 8 \\ 
   \hline
   $|2;2;1\rangle$ & $|2;2;-1\rangle$ & $|0;0;0\rangle$ & $|0;0;0\rangle$ & $2^{-\beta^2}\left(1/2+\frac{\beta^2}{32}\right)^{2}$ & 8 \\ 
   \hline
   $|1;1;1\rangle$ & $|0;0;0\rangle$ & $|1;1;-1\rangle$ & $|0;0;0\rangle$ & $2^{-2\beta^2}\left(1/4\right)^{2}$ & 4 \\ 
   \hline
   $|2;2;1\rangle$ & $|0;0;0\rangle$ & $|2;2;-1\rangle$ & $|0;0;0\rangle$ & $2^{-2\beta^2}\left(1/16+\frac{\beta^2}{8}\right)^{2}$ & 4 \\ 
   \hline
   $|2;2;1\rangle$ & $|1;1;-1\rangle$ & $|0;0;0\rangle$ & $|0;0;0\rangle$ & $2^{-\beta^2}\left(-\frac{1}{2\sqrt{2}}+\frac{\beta^2}{8\sqrt{2}}\right)^{2}$ & 16 \\ 
   \hline
   $|1,1;1,1;1\rangle$ & $|0;0;-1\rangle$ & $|0;0;0\rangle$ & $|0;0;0\rangle$ & $2^{-\beta^2}\frac{1}{2}\left(1/4\right)^{2}$ & 16 \\ 
   \hline
   $|1,1;1,1;1\rangle$ & $|0;0;0\rangle$ & $|0;0;-1\rangle$ & $|0;0;0\rangle$ & $2^{-2\beta^2}\frac{1}{2}\left(1/4-\frac{\beta^2}{4}\right)^{2}$ & 8 \\ 
   \hline
   $|1,1;1,1;0\rangle$ & $|0;0;0\rangle$ & $|0;0;0\rangle$ & $|0;0;0\rangle$ & $\frac{1}{2}\left(1/4\right)^{2}$ & 4 \\ 
   \hline
   $|1,1;2;1\rangle$ & $|0;0;-1\rangle$ & $|0;0;0\rangle$ & $|0;0;0\rangle$ & $2^{-\beta^2}\frac{1}{4\sqrt{2}}\frac{\beta}{4\sqrt{2}}$ & 16 \\ 
   \hline
   $|1,1;2;-1\rangle$ & $|0;0;1\rangle$ & $|0;0;0\rangle$ & $|0;0;0\rangle$ & $-2^{-\beta^2}\frac{1}{4\sqrt{2}}\frac{\beta}{4\sqrt{2}}$ & 16 \\ 
   \hline
   $|1,1;2;1\rangle$ & $|0;0;0\rangle$ & $|0;0;-1\rangle$ & $|0;0;0\rangle$ & $2^{-2\beta^2}\frac{1}{2\sqrt{2}}\frac{\beta}{4\sqrt{2}}$ & 8 \\ 
   \hline
   $|1,1;2;-1\rangle$ & $|0;0;0\rangle$ & $|0;0;1\rangle$ & $|0;0;0\rangle$ & $-2^{-2\beta^2}\frac{1}{2\sqrt{2}}\frac{\beta}{4\sqrt{2}}$ & 8 \\ 
   \hline
   \hline
\end{tabular}
\caption{Here we report the different generalized Renyi entropies, $R_{1,2;3,4}$, that determine the leading corrections to the $R_2(t)$ at early times.  The multiplicity indicates the number of related generalized Renyi entropies (obtained through permuting the order of states in the entropy) that equal $R_{1,2;3,4}$.  So for example the first entry in the table $R_{0,0,0,0}^{1,-1,0,0}$ equals $R_{0,0,0,0}^{-1,1,0,0}, R_{0,0,0,0}^{1,0,0,-1}, R_{0,0,0,0}^{-1,0,0,1},R_{0,0,0,0}^{0,0,1,-1},R_{0,0,0,0}^{0,0,-1,1},R_{0,0,0,0}^{0,1,-1,0}$, and $R_{0,0,0,0}^{0,-1,1,0}$ (adapting the notation of Eqn. S1. 5) for a multiplicity of 8.  }
\label{statenumber}
\end{table*}

While we do not report here general formulae for the GMSREs needed to compute the third Renyi entropy, $R_3(t)$, (these are considerably more involved as they involve computing 6-point conformal correlation functions on a 3-sheeted Riemann surface), we can compute the handful of GMSREs needed to compute $R_3(t)$ at early times.  Doing so for the sextuplets involving $\ket{0,0}, \ket{n,\pm 1}, n=0,1$, we find
\begin{equation}
R_3(t)=L^2 J_1^2 (\frac{2\pi}{L})^{2\beta^2} t^2 f_3(\beta),
\end{equation}
where $f_3(\beta)$ for small $\beta$ reads
\begin{equation}
f_3(\beta)\simeq\left(-\frac{31}{27} +3 \log \frac{27}{16} \log 3 \right) \beta^4.
\end{equation}
We again see the $\beta^4t^2$ dependence and again find that we cannot ignore the contribution of states with non-trivial bosonic mode content.

Like with $R_2(t)$ and $R_3(t)$, we can use unitary perturbation theory to compute the time dependence of the order parameter, $C(t)=\langle \cos(\beta\phi)\rangle(t)$.  Here we find that the contribution of states involving only vertex operators, $\ket{0;0;m=0,\pm 1}$, determine $C(t)$ at leading order in $\beta$ and $t$ to be: 
\begin{equation}
    C(t) = -(2\pi)^{1+2\beta^2}J_1 \beta^2 t^2L^{-2\beta^2}.
\end{equation}
We see that this contribution comes in at ${\cal O}(\beta^2)$.  At early times $C(t)$ is then determined solely by the dynamics of the compact zero mode of the field, i.e. the problem is quantum mechanical not quantum field theoretic.

\section{Thermal Asympototics of the Post-Quench System}
\label{SI:TBA}
In the quantum quench in which we are interested, the system is initially prepared in the ground state of a Luttinger liquid with $J_1=0$ and is allowed to evolve with the finite $J_1$ sine-Gordon Hamiltonian $H$, of Eqn. 3 in the main text.  In this protocol, the entire system is always in a pure state, but the reduced density matrix of an arbitrary finite compact subsystem attains a long time limit that can be described by a statistical ensemble and where at asymptotically long times, all local physical observables relax to stationary values.   

For a generic system, the properties of its reduced density matrix are captured by a Gibbs (thermal) ensemble.  However for an integrable model, like the sine-Gordon at hand, the appropriate ensemble is a generalised Gibbs ensemble (GGE) where the ensemble accounts for the higher conserved charges, $Q_i$, present in integrable systems.  It is an interesting question however how close the GGE here is to a thermal ensemble, or equivalently, whether the generalized temperatures, $T_i$, corresponding to the higher charges are close to $\infty$.  We can answer this question quantitatively for the two quantities that we have measured in the quench, $R_2(t)$ and $C(t)$.   We know how much energy, $Le$, that we have injected into the system where $e$ is given by 
\begin{equation}\label{eq:betatilde}
e = -\frac{\pi}{6L^2} + M_s^2\tan(\pi\xi/2)/4.
\end{equation}
If the quench were to be described by a thermal ensemble, this energy density would be associated with a temperature $T=\tilde\beta^{-1}$ (we use $\tilde\beta$ to distinguish the inverse temperature from the sine-Gordon coupling, $\beta$).  Using the thermodynamic Bethe ansatz (TBA), we can connect this energy density $e$ with an effective temperature $T$.  Once we know this temperature, we can, again using the TBA, then compute what the asymptotic values of $R_2(t)$ and $C(t)$ would be if the late time dynamics were to be described by the thermal ensemble at this temperature.  As we showed in the main text, there are considerable differences between the measured values of $R_2(t)$ and $C(t)$ using the TSM approach and these putative thermal values.  This indicates that the quench dynamics are far from being thermal and that the constraints introduced by the conservation of higher conserved charges are crucial for understanding the long time asymptotics.

\subsection{Thermodynamic Bethe Ansatz Equations for the Sine-Gordon Model at Its Reflectionless Points}

The TBA equations provide expressions for the energy and the free energy of the sine-Gordon model.  These equations are relatively simple when the scattering of the theory is diagonal.  This occurs when the parameter $\xi^{-1}$,
\begin{equation}
\xi^{-1} = \frac{2}{\beta^2} -1,
\end{equation}
is an integer.  As $\beta\rightarrow 0$, the values of $\beta$ that corresponds to reflectionless points becomes dense.  We generically expect that physical quantities like $R_2(t=\infty)$ and $C(t=\infty)$ that are connected to the free energy of the system will depend smoothly on $\beta$.  Thus even for those values of $\beta$ where non-diagonal scattering is present, we expect to be able to use the reflectionless TBA equations at the closest integer $\xi^{-1}$ to compute $R_2(t=\infty)$ and $C(t=\infty)$.  Thus for $\beta=3/20$, we will use $\xi^{-1}=87$ to derive the associated values of $R_2$ and $C$. For the other two values of $\beta$ considered in the main text, $\beta=1/\sqrt{8}$ and $\beta=1/\sqrt{2}$, we do not need to make this approximation as the associated values of $\xi$ are integer-valued as is.

The basic ingredient of the TBA equations are the S-matrices of the fundamental excitations of the model.
The excitations of sine-Gordon consists of $\xi^{-1}+1$ particles.  The first $\xi^{-1}-1$ particles are breathers (labeled as $n=1,\cdots,\xi^{-1}-1$) while the last two particles are the soliton and anti-soliton, $n=\xi^{-1},\xi^{-1}+1\equiv\pm$.  The S-matrices for these particles are as follows: \cite{rtv-93,km-90}
\begin{eqnarray}
S_{+-}(\theta) &=& S_{++}(\theta)(-1)^{\xi^{-1}+1}=\prod^{\xi^{-1}-1}_{k=1}f_{k\xi}(\theta);\cr\cr
S_{\pm n}(\theta) &=&  \prod F_{1/2-\frac{n-2k}{2\xi^{-1}}}(\theta), ~~~n=1,\cdots,\xi^{-1}-1;\cr\cr
S_{nm}(\theta) &=& F_{\frac{|n-m|}{2\xi^{-1}}}(\theta) \bigg[\prod^{{\rm min}(n,m)-1}_{k=1}F_{\frac{|n-m|+2k}{2\xi^{-1}}}(\theta)\bigg]^2 F_{\frac{n+m}{2\xi^{-1}}}(\theta), ~~~ n,m\leq \xi^{-1}-1;\cr\cr
f_\alpha(\theta) &=& \frac{s((\theta+i\alpha\pi)/2)}{s((\theta-i\alpha\pi)/2)};\cr\cr
F_\alpha(\theta) &=& f_\alpha(\theta)f_\alpha(i\pi-\theta);\cr\cr
s(\theta)/c(\theta) &\equiv& \sinh(\theta)/\cosh(\theta).
\end{eqnarray}
The parameter $\theta$ here is a rapidity that governs the energy/momentum of an excitation of mass $m$: $m\cosh(\theta)/m\sinh(\theta)$.

With these S-matrices in hand, one can straightforwardly write down an expression for free energy density, $f(\tilde\beta)$:
\begin{eqnarray}
f(\tilde\beta) &=& -\frac{1}{\tilde\beta} \sum^{\xi^{-1}+1}_{n=1} m_n\int^\infty_{-\infty} \frac{d\theta}{2\pi} c(\theta) L_{-n}(\theta),
\end{eqnarray}
where the mass $m_n$ of the excitations are
\begin{eqnarray}
m_n &=& 2m_s\sin((\pi n \xi/2)), ~~~ n=1,\cdots \xi^{-1}-1;\cr\cr
m_\pm &\equiv& m_s= (2J_1)^{(2-\beta^2)^{-1}}
\frac{2\Gamma(\xi/2)}{\sqrt{\pi}\Gamma(1/2+\xi/2)}\big(\frac{\pi\Gamma(1-\beta^2/2)}{2\Gamma(\beta^2/2)}\big)^{1/(2-\beta^2)},
\end{eqnarray}
while the functions, $L_{\pm n}$, are defined by
\begin{eqnarray}
L_{\pm n}(\theta) &\equiv& \log(1+e^{\pm\epsilon_n(\theta)}).
\end{eqnarray}
Here $\epsilon_n$ are so-called pseudoenergies and are defined by the set of coupled equations
\begin{eqnarray}
\epsilon_{n0}(\theta) &=& \epsilon_n(\theta) + \sum^{\xi^{-1}+1}_{k=1}\int^\infty_{-\infty} \frac{d\theta'}{2\pi} \phi_{nk}(\theta-\theta')L_{-k}(\theta');\cr\cr
\epsilon_{n0}(\theta) &\equiv& m_n\tilde\beta\cosh(\theta).
\end{eqnarray}
Finally the $\phi_{ab}$ are kernels derived from the S-matrices above and are defined in terms of the logarithmic derivative of $S_{ab}$:
\begin{eqnarray}
\phi_{ab}(\theta) &=& -i\partial_\theta \log S_{ab}(\theta);\cr\cr
\phi_\alpha(\theta) &\equiv& -i\partial_\theta \log f_\alpha(\theta) = -\frac{\sin(\pi\alpha)}{c(\theta)-\cos(\alpha\pi)}.
\end{eqnarray}
The last identity in the above is useful for writing down the log-derivatives of the various $S_{ab}$.

The energy density, $e(\tilde\beta)$, is defined in terms of the $L_{-n}$'s as well:
\begin{eqnarray}
e(\tilde\beta) &=& \frac{1}{\tilde\beta}m_n\sum^{\xi^{-1}+1}_{n=1}\int^\infty_{-\infty} \frac{d\theta}{2\pi} c(\theta) L_{-n}(\theta)\cr\cr
&& +  \sum^{\xi^{-1}+1}_{n=1} m_n\int^\infty_{-\infty} \frac{d\theta}{2\pi} c(\theta) \partial_{\tilde\beta} \epsilon_n(\theta) \frac{e^{-\epsilon_n(\theta)}}{1 + e^{-\epsilon_{n}(\theta)}}.
\end{eqnarray}
To find the temperature $\tilde\beta$ that would correspond to our quench if the quench was thermal, we solve the following equation for $\tilde\beta$
\begin{equation}\label{eq:betatilde}
e(\tilde\beta) = -\frac{\pi}{6L^2} + M_s^2\tan(\pi\xi/2)/4.
\end{equation}
Finally we can write down the associated thermal densities of the excitations as a function of rapidity,
\begin{eqnarray}
\rho_a(\theta) &=&\rho_{a0}(\theta)+\sum^{\xi^{-1}+1}_{n=1} \int^\infty_{-\infty} \frac{d\theta'}{2\pi} \phi_{ab}(\theta-\theta')\frac{\rho_b(\theta')}{1 + e^{\epsilon_{b}(\theta')}},
\end{eqnarray}
where $\rho_{a0}(\theta) =\frac{m_a}{2\pi}\cosh(\theta)$ is the bare density (the density absent any interactions in the system).

These equations for the pseudoenergies, $\epsilon_n$, can be recast into a universal form in terms of the incidence matrix of the $D_{2\xi^{-1} + 2}$ Dynkin diagram (the Dynkin diagram for $SO(2(\xi^{-1}+1)))$.
We can write
\begin{eqnarray}\label{dynkinTBA}
\epsilon_{a}(\theta) &=& \epsilon_{a0}(\theta) - \sum_b G_{ab}\int^\infty_{-\infty} \frac{d\theta'}{2\pi} \phi_\xi(\theta-\theta')(\epsilon_{b0}(\theta')-L_{+b}(\epsilon_{b}(\theta')));\cr\cr
\phi_\xi(\theta) &=& \frac{h}{2\cosh(h\theta/2)};\cr\cr
h &=& 2(\xi^{-1}+1)-2,
\end{eqnarray}
where $G_{ab}$ is the incidence matrix for the $D_{2\xi^{-1}+2}$ Dynkin diagram and $h$ is the corresponding dual Coexter number for the algebra. $G_{ab}$ is defined such that if there is a bond between nodes $a$ and $b$ of the diagram, then $G_{ab}=1$,  otherwise $G_{ab}=0$.  These equations do not admit analytic solutions, but can be solved through iteration, by taking as an initial ansatz $\epsilon_a=\epsilon_{a0}$, substituting this ansatz into the integrals on the r.h.s. of the first equation in Eqn.\ref{dynkinTBA} so finding a new value of $\epsilon_a$, and then repeating the process up to convergence of the solution. 
In this universal formulation, the density of states reads
\begin{eqnarray}
\rho_{a}(\theta) &=& \rho_{a0}(\theta) + \sum_b G_{ab}\int^\infty_{-\infty} \frac{d\theta'}{2\pi} \phi_\xi(\theta-\theta')\left(\frac{\rho_b(\theta')}{1+e^{-\epsilon_b(\theta')}} -\rho_{b0}(\theta')\right).
\end{eqnarray}
From $\rho_{a}(\theta)$, one can derive the average occupancy per unit length, $N_a$, and absolute velocity, $v_a$, for each of the excitations:
\begin{eqnarray}
N_a&=& \int^\infty_{-\infty} d\theta \frac{\rho_a(\theta)}{1+e^{\epsilon_a(\theta)}} ;\cr\cr
v_a &= & \frac{1}{N_a}\int^\infty_{-\infty} d\theta |\tanh(\theta)|\frac{\rho_a(\theta)}{1+e^{\epsilon_a(\theta)}} .
\end{eqnarray}
These quantities allow one to understand both which excitations are created in the course of the quench and how close to the `speed of light' they are moving on average.

With the free energy in hand and the effective temperature $\tilde\beta$ known, the n-th thermal R\'enyi entropy has a simple expression in terms of the free energies at $\tilde\beta$ and $n\tilde\beta$:
\begin{equation}
R_n(t)=\frac{1}{1-n}\left[\log \mathrm{Tr}e^{-n\tilde{\beta} H(t)}-n\log  \mathrm{Tr}e^{-\tilde{\beta} H(t)} \right],
\end{equation}
where $\log \mathrm{Tr}e^{-n\tilde{\beta} H(t)}$ is the free energy of a system with inverse temperature $n\tilde{\beta}\equiv n/T$ while $\log \mathrm{Tr}e^{-\tilde{\beta} H(t)}$ is the free energy with inverse temperature $\tilde{\beta}$.

As with $R_2$, we can also compute the thermal value of the order parameter $\cos(\beta \phi)(t)$ using the free energy: 
\begin{equation}
C(\infty) =\cos(\beta \phi)(t=\infty)=-\partial_{J_1}f(\tilde{\beta}).
\end{equation}
In Table 1 of the main text we report the thermal values of $R_2$ and $C$. As we have said, they are not predictive for our problem as the thermal ensemble turns out to be far away from the GGE describing the quench. 

\section{Truncated Spectrum Methods}
\label{SI:TSM}
\subsection{Basics of the Approach}
Truncated spectrum methods (TSMs) were developed in two papers by V. Yurov and Al. Zamolodchikov, one treating perturbations of the scaling Yang-Lee model~\cite{yurov1990truncated}, and one treating the critical Ising model perturbed by a magnetic field~\cite{yurov1991truncated}.
In both cases, the basic formulation of the problem is the same.  The TSM enables the study of a Hamiltonian of the following form:
\begin{equation}
H = H_{\rm known} + J_1 V_{\rm pert}.
\end{equation}
For our purposes, $H_{\rm known}$ is a $c=1$ compact boson, and $V_{\rm pert}$ involves the perturbing cosine operator of the sine-Gordon model, 
$$
V=\int^L_0\cos(\beta\Phi(x)).
$$
Here $L$ is the volume of the system.  A key element of the method is that we work in finite volume.

The space of eigenstates of the $c=1$ boson, that of $H_{\rm known}$, is employed by the TSM as a computational basis.  Because $L$ is finite, this spectrum is discrete.  This spectrum can be understood by considering the mode expansion of the boson~\cite{CFTBook} 
\begin{eqnarray}
\Phi (x,t) &=& \Phi_0 + \frac{4\pi}{L}\Pi_0 t + \frac{2\pi m}{\beta L}x
+ i\sum_{l\neq 0} \frac{1}{l}\Big(a_l e^{\frac{2\pi i l}{L}(x-t)}-\bar a_{-l}e^{\frac{2\pi i l}{L}(x+t)} \Big). \qquad
\end{eqnarray}
This mode expansion assumes the boson has compactification radius $2\pi/\beta$, i.e. $\Phi(x+L,t) = \Phi(x,t) + \frac{2\pi}{\beta} m$, 
where $m$ denotes the winding number, which is related to the $U(1)$ charge of the sector. The operator $\Phi_0$ 
is the `center of mass' of the Bose field and $\Pi_0$ is its conjugate momentum, which has permitted values $n\beta$, with integer $n$. 
These obey the commutator $[\Phi_0,\Pi_0] = i$.

The bosonic Hilbert space emerges from an infinite set of highest weight states marked by the bosonic winding number and the value of conjugate momentum:
\begin{eqnarray}
|n,m\rangle = e^{in\beta \Phi(0) + i\frac{m}{2\beta} \Theta(0)}|0\rangle.
\end{eqnarray}
These highest weight states $|n,m\rangle$ are defined by acting with vertex operators involving the boson and its dual on the vacuum $|0\rangle$. The dual boson, $\Theta$, can be defined via the relation
\begin{equation}
\partial_x\Theta(x,t) = \partial_t \Phi (x,t).
\end{equation}
The quantum number $n$ gives the momentum of the bosonic zero mode for the state while the quantum number $m$ gives the $U(1)$ charge of the state.

The full Hilbert space is then constructed by the acting with the right and left moving modes ($a_n$ and $\bar a_n$) of the field on the highest weight states:
\begin{equation}\label{bosonic_states}
|\Psi\rangle = \prod^M_{j=1}a_{k_j}\prod^{\bar M}_{\bar j=1}\bar a_{k_{\bar j}}|n,m\rangle.
\end{equation}
The energy and momentum of such a state is
\begin{eqnarray}
E_\Psi &=& \frac{2\pi}{R}\bigg(n^2\beta^2 + \frac{m^2}{4\beta^2}+\sum^M_{j=1} k_j + \sum^{\bar M}_{\bar{j}=1} k_{\bar j}  -\frac{1}{12}\bigg),\cr\cr
P_\psi &=& \frac{2\pi}{R}\bigg((n-m) + \sum^M_{j=1} k_j - \sum^{\bar M}_{\bar{j}=1} k_{\bar j}\bigg). 
\end{eqnarray}
The $1/12$ term in $E_\Psi$ reflects the fact that the vacuum energy in the conformal limit on the cylinder does not vanish if it is assumed to be zero on the plane. The $a_n/\bar a_{n}$ satisfy the following commutation relations:
\begin{eqnarray}
[ a_n,a_{m}] &=& n\delta_{n+m,0};\cr\cr
[ \bar a_n,\bar a_{m}] &=& n\delta_{n+m,0};\cr\cr
[ a_n,\bar a_{m}] &=& 0.
\end{eqnarray}
These commutators, together with the relation governing commuting the modes with vertex operators
\begin{equation}
[ a_n, e^{i\beta \Phi(0)} ] = -\beta e^{i\beta \Phi(0)},
\end{equation}
allow one to compute generic matrix elements of the states with the vertex operators appearing in the sine-Gordon Hamiltonian.  

Using our ability to compute matrix elements of $V_{\rm pert}$, we can represent the full sine-Gordon Hamiltonian in matrix form:
\begin{equation}
H = \begin{bmatrix}
E_1 +J_1\langle E_1|V_{\rm pert}|E_1\rangle  & J_1\langle E_1|V_{\rm pert}|E_2\rangle  & J_1\langle E_1|V_{\rm pert}|E_3\rangle  & \dots   \\
J_1\langle E_2|V_{\rm pert}|E_1\rangle  &  E_2 +J_1\langle E_2|V_{\rm pert}|E_2\rangle  & J_1\langle E_2|V_{\rm pert}|E_3\rangle & \dots  \\
J_1\langle E_3|V_{\rm pert}|E_1\rangle  &  J_1\langle E_3|V_{\rm pert}|E_2\rangle  & E_3 + J_1\langle E_3|V_{\rm pert}|E_3\rangle & \dots  \\
\vdots & \vdots & \vdots & \ddots \\
\end{bmatrix}.
\end{equation}
In this form $H$ is an infinite dimensional matrix.  Here we will truncate the spectrum, keeping only the first $N$ states.  This leaves the Hamiltonian matrix, $H_N$, as finite dimensional:

\begin{equation}
H_N = \begin{bmatrix}
E_1 +J_1\langle E_1|V_{\rm pert}|E_1\rangle  & J_1\langle E_1|V_{\rm pert}|E_2\rangle  & \dots & J_1\langle E_1|V_{\rm pert}|E_N\rangle  \\
J_1\langle E_2|V_{\rm pert}|E_1\rangle  &  E_2 +J_1\langle E_2|V_{\rm pert}|E_2\rangle & \dots & J_1\langle E_2|V_{\rm pert}|E_N\rangle \\
\vdots & \vdots & \ddots \\
J_1\langle E_N|V_{\rm pert}|E_1\rangle  &  J_1\langle E_N|V_{\rm pert}|E_2\rangle  & \dots & E_N + J_1\langle E_N|V_{\rm pert}|E_N\rangle \\
\end{bmatrix}
\end{equation}

To analyze the properties of the model, we then numerically diagonalize the matrix, obtaining information on its spectrum and matrix elements.

\begin{table*}[t]
\centering
\begin{tabular}{||c|c|c|c|c|c|c||} 
 \hline
 $\beta$ & L & $J_1$ & $N_{zm}$ & $RE_{\psi,osc}/2\pi$ & $N_{total,symm.-red}$ & $N_{total,no-symm.}$  \\ 
 \hline \hline
  $3/20$ & 20 & 0.4 & 163 & 5 & 4205 & 13789 \\ 
  \hline
  $3/20$ & 20 & 0.4 & 163 & 6 & 9257 & 32242 \\ 
  \hline
  $3/20$ & 20 & 0.4 & 163 & 7 & 17361 & 62625 \\ 
  \hline
  $3/20$ & 20 & 0.4 & 163 & 8 & 37549 & 139399 \\ 
  \hline
  $3/20$ & 50 & 0.065 & 143 & 5 & 3665 & 12009 \\ 
  \hline
  $3/20$ & 50 & 0.065 & 143 & 6 & 8057 & 28042 \\ 
  \hline
  $3/20$ & 50 & 0.065 & 143 & 7 & 16644 & 60023 \\ 
  \hline
  $3/20$ & 50 & 0.065 & 143 & 8 & 32499 & 120569 \\ 
  \hline
  $1/\sqrt{8}$ & 30 & 0.375 & 43 & 5 & 983 & 3167 \\ 
  \hline
  $1/\sqrt{8}$ & 30 & 0.375 & 43 & 6 & 2141 & 7342 \\ 
  \hline
  $1/\sqrt{8}$ & 30 & 0.375 & 43 & 7 & 4125 & 14655 \\ 
  \hline  
  $1/\sqrt{8}$ & 30 & 0.375 & 43 & 8 & 8306 & 30387 \\ 
  \hline
  $1/\sqrt{2}$ & 20 & 0.375 & 21 & 5 & 492 & 1553 \\
   \hline
  $1/\sqrt{2}$ & 20 & 0.375 & 21 & 6 & 1056 & 3550 \\
   \hline
  $1/\sqrt{2}$ & 20 & 0.375 & 21 & 7 & 2080 & 7263 \\
   \hline  
  $1/\sqrt{2}$ & 20 & 0.375 & 21 & 8 & 4237 & 15251 \\
   \hline
\end{tabular}
\caption{Here we report for the three values of $\beta$ considered herein the number of zero mode states, $N_{zm}$, in the simulation and the total number of states used in the simulations at different values of $E_{\psi,osc}$.  We report both the symmetry-reduced number of states, $N_{total,symm.-red}$, as well as the number of states, $N_{total,no-symm.}$, that would be present (approximately) in the simulation absent the application of symmetry.}
\label{statenumber}
\end{table*}

\subsection{Implementation of Symmetries and the TSM Cutoff}
\label{subsec:cutoff}
In this subsection, we describe how to choose our finite basis of states for performing the computation.  We first reduce this computational basis by invoking symmetries.  Our quench is going to take place in a sector of the theory with 0 U(1) charge, even under the $Z_2$ symmetry $\Phi \rightarrow -\Phi$, even under parity, $\Phi(x)\rightarrow \Phi(-x)$, and for which the momentum, $P_\psi$ is zero.  This means our computational basis will consist of states of the form:
\begin{eqnarray}
    |\Psi\rangle &=& \prod^M_{j=1}a_{k_j}\prod^{\bar M}_{\bar j=1}\bar a_{k_{\bar j}}|n,0\rangle + \prod^M_{j=1}\bar a_{k_j}\prod^{\bar M}_{\bar j=1} a_{k_{\bar j}}|n,0\rangle \cr\cr 
    &+& (-1)^{M+\bar M}\prod^M_{j=1}a_{k_j}\prod^{\bar M}_{\bar j=1}\bar a_{k_{\bar j}}|-n,0\rangle + 
    (-1)^{M+\bar M}\prod^M_{j=1}\bar a_{k_j}\prod^{\bar M}_{\bar j=1} a_{k_{\bar j}}|-n,0\rangle.
\end{eqnarray}
This still leaves, however, an infinite set of states.  Typically in TSM studies one, as a first approximation, truncates the states in energy, i.e. one excludes all states whose energy, $E_\Psi$, exceeds some cutoff, $E_{cutoff}$.  Here we modify this approach.  We will treat the contribution to the energy coming from the highest weight part of the state, i.e., the zero mode contribution,  
$$
E_{\Psi,zero-mode}=\frac{2\pi}{R}n^2\beta^2,
$$
differently from the oscillator contribution to a state's energy:
$$
E_{\Psi,osc.} = \frac{2\pi}{R}\bigg(\sum^M_{j=1} k_j + \sum^{\bar M}_{\bar{j}=1} k_{\bar j}\bigg).
$$
We correspondingly introduce two cutoffs, $E_{c}$ and $E_{c,osc.}$.  Our finite computational basis will then be formed of states which satisfy:
\begin{equation}
    E_{\Psi,zero-mode} + E_{\Psi,osc.} < E_{c}; ~~~~~ E_{\Psi,osc.}/2 < E_{c,osc.},
\end{equation}
where with the factor of $1/2$ in the above, we are defining $E_{c,osc.}$ in terms of the energy of the chiral part of the state (because we work in a zero momentum sector, the energies of the chiral and anti-chiral parts of the state are always equal).
The rational for this choice is based on the observation that much of the physics for our quench is determined by the dynamics of the zero mode, particularly for small values of $\beta$.  It thus made sense to choose a much larger cutoff for a state's energy as a whole, $E_{c}$, than the cutoff applied to the oscillator part of a state's energy, $E_{c,osc.}$.  In practice, to determine $E_{c}$, we first studied the model absent any oscillator modes (i.e. $E_{c,osc.}=0$).  We then chose $E_{c}$ sufficiently large that convergence in the quench dynamics was obtained (i.e. further increases in $E_{c}$ led to no changes in the results).  Having determined $E_{c}$, we then systematically increased $E_{c,osc.}$ from zero, studying its effect on the results.  When we could not always increase $E_{c,osc.}$ to the point of convergence (i.e. again, the results were completely unchanging), we developed an extrapolation procedure for our data.  This is described in Section \ref{SI:Extrapolation}.  In the Tab. \ref{statenumber}, we provide a table giving the number of states for some of the different simulations.  In general we found that at small $\beta$ we needed to include many more zero-mode states in the simulation for convergence, leading to the need to deal with much large Hilbert spaces.  We also see that taking into account basic $Z_2$ symmetries reduces the Hilbert space by a factor of 4.

\subsection{TSM for Non-Equilibrium Studies}
\label{sec:tsm}
\begin{table}[h!]
\centering
\begin{tabular}{||c|c|c|c|c|c|c||} 
 \hline
 $\beta$ & $\tilde\beta$ & $J_1$ & L & Particle type, a & $v_{a,avg}$ & $N_a$ \\ 
 \hline \hline
  3/20 & 1.134 & 0.4 & 20 & 1& 0.778 & 0.136\\ 
  \hline
  3/20 & 1.134 & 0.4 & 20  & 2 & 0.672 & 0.075 \\ 
  \hline
  3/20 & 1.134 & 0.4 & 20  & 3 & 0.595 & 0.041 \\ 
  \hline
  3/20 & 1.134 & 0.4 & 20  & 4 & 0.486 & 0.026 \\ 
  \hline
  3/20 & 1.134 & 0.4 & 20  & 5 & 0.447 & 0.017 \\ 
  \hline
  3/20 & 1.134 & 0.4 & 20  & 6 & 0.414 & 0.011 \\ 
 \hline\hline
  $1/\sqrt{8}$ & 3.477 & 0.0375 & 30& 1& 0.632 & 0.036\\ 
  \hline
  $1/\sqrt{8}$  & 3.477 & 0.0375 & 30& 2& 0.5 & 0.015 \\ 
 \hline\hline
   $1/\sqrt{2}$ & 4.220 & 0.0375 & 20& 1,3,4& 0.477& 0.013 \\ 
  \hline
  $1/\sqrt{2}$  & 4.220 & 0.0375 & 20 & 2& 0.373 & 0.003\\ 
  \hline\hline
\end{tabular}
\caption{Here we report the values of various parameters associated with the thermal values of the different particle types as determined from the TBA analysis for the different quenches considered here.}
\label{TAB:TBA}
\end{table}

In this section we explain how we compute non-equilibrium quench dynamics using TSM.  Our quench amounts to studying how the ground state of the $J_1=0$ system (that of of a $c=1$ boson) evolves after a finite coupling $J_1$ is turned on at $t=0$.  This is a particularly simple quench for us to study as the state at $t=0$, $|\Phi(t=0)\rangle$, is a state in our computational basis.   To compute the time evolution of the state, we use the TSM to compute the spectrum of the post-quench Hamiltonian:
\begin{equation}
    H(J_1)|E_n\rangle = E_n|E_n\rangle, ~~~ {n=1,\cdots, N}.
\end{equation}
These eigenstates are expressed by the TSM in our computational basis:
\begin{equation}
    |E_n\rangle = \sum_{i=1}^N c_{ni} |\Psi_i\rangle.
\end{equation}
The first state in this basis, $|\Psi_1\rangle \equiv |0\rangle$, is our state at $t=0$.  Thus the time evolution of the state $|\Phi(t)\rangle$ is given by
\begin{eqnarray}
    |\Phi(t)\rangle &=& \sum_{n=1}^N e^{iE_nt}c^*_{n1}|E_n\rangle\cr\cr
    &=& \sum^N_{n=1,i=1}e^{iE_nt} c^*_{n1} c_{ni} |\Psi_i\rangle\cr\cr
    &=& \sum^N_{i=1}\alpha_i(t)|\Psi_i\rangle,
\end{eqnarray}
where in the last line we have expressed the time evolved state as a linear combination of time-dependent coefficients in our computational basis.

In computing time dependent properties, there are two questions in regards to the interpretation of the data.  1) What is the dependence of the data on the two cutoffs, $E_{c,zero-mode}$ and $E_{c,osc.}$? 2) What is the dependence on the system size?

To address the first question, we have chosen $E_{c,zero-mode}$ to be large enough that the data is effectively converged (at approximately the $10^{-4}$ level) at a given $E_{c,osc.}$ for the times out to which the simulation was run.  We could not obtain a similar level of convergence by choosing $E_{c,osc.}$ sufficiently large. In varying $E_{c,osc.}$, we were still seeing corresponding variations in time-dependent quantities on the order of $10^{-2}$.  Thus we pursued an extrapolation strategy to extrapolate the data to $E_{c,osc.}=\infty$.  This is described in Section \ref{SI:Extrapolation}.

To answer the second question, what is the dependence on volume, one has to have some understanding of the energy injected into the system.  This energy goes into the creation of pairs of quasi-particles with some characteristic velocity, $v$.  Because the system has a Lorentz symmetry, $v<c(=1)$.  For times $t<L/2v$ ($L/2v$ is the time needed for a pair of counterpropagating quasi-particles to traverse the system and meet up again), the dynamics will appear as if in infinite volume.  At times $t>L/2v$, the system will become realize that it is in fact of finite size.  We can estimate $v$ as follows.  The TBA of Section \ref{SI:TBA} allows us to estimate the average {\it thermal} velocity of each of the different types of quasi-particles.  We present these in Table \ref{TAB:TBA}.
While the long time behaviour of the quench is not governed by a thermal density matrix, these velocities give us an idea of what the average velocity of the post-quench system's quasi-particle are.

\begin{figure}[t]
	\includegraphics[width=0.5\textwidth]{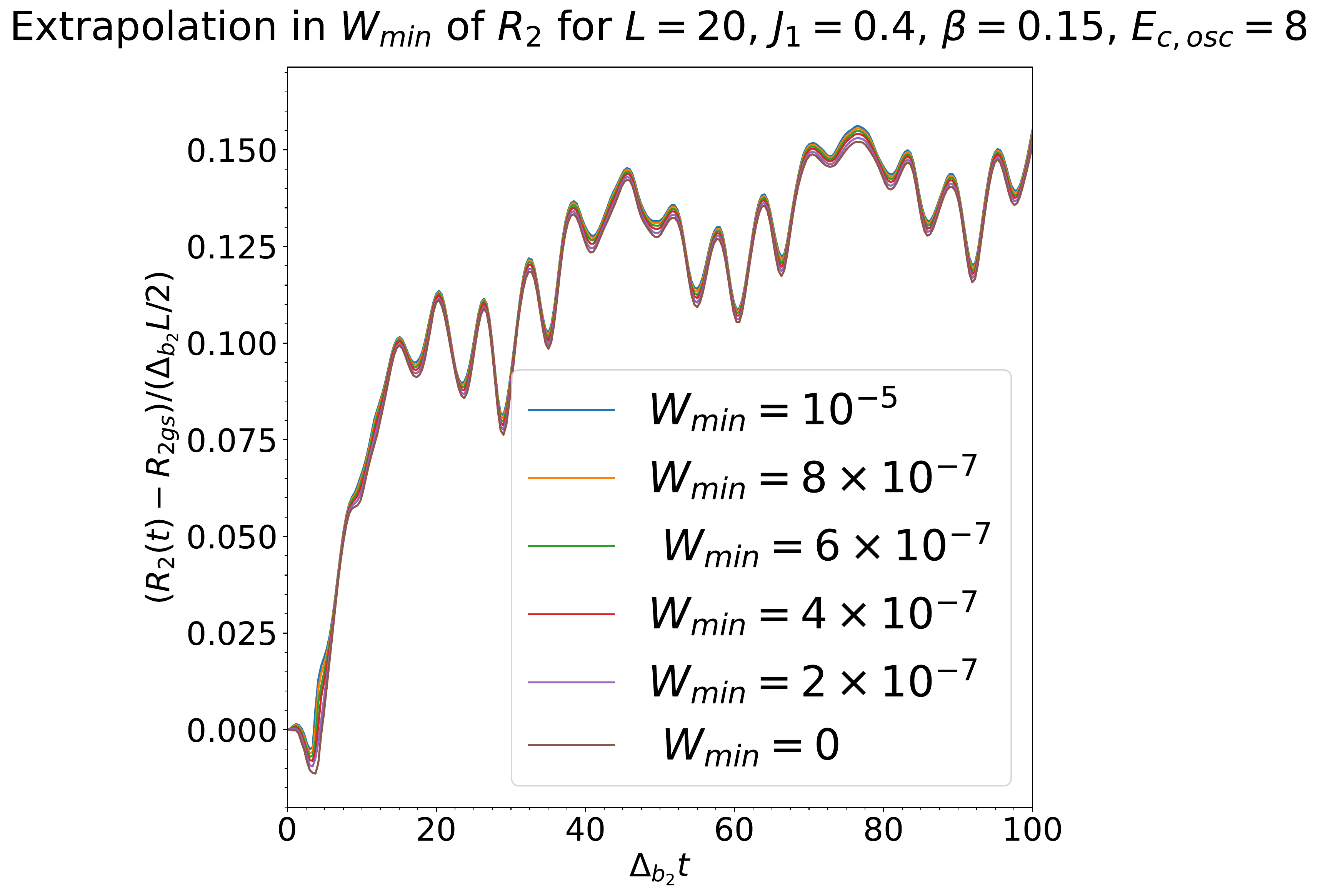}
	\caption{Here we extrapolate the Renyi entropy computed at fixed $L=20$ and $E_c$ but at different $W$'s to $W=0$.} \label{fig:REW}
\end{figure}

To compute the time evolution of states within TSMs, a different option for time evolution is to expand the time evolution operator in terms of Chebyshev polynomials, a method developed in Ref. \onlinecite{cheb_timeev} for quantum quenches in the Ising model and used successful in this context in Refs. \onlinecite{jkr,rjk} . While we did not benchmark this approach against the approach used here, it would be interesting to understand which methodology is preferable in accessing longer times in the context of sine-Gordon quenches at small $\beta$.

\subsection{Extrapolation of Data}

\label{SI:Extrapolation}
The extrapolation in $W_{min}$ has been done by fixing $E_{c,osc}$ and using as an extrapolation form
\begin{equation}
R_{2}(t)-R_{2,gs}=a(t)+c_2(t)W_{min}^{\gamma(t)}, 
\end{equation}
in order to extract $a(t)$, i.e. the second R\'enyi entropies at $W_{min}=0$. An example of this extrapolation is given in Fig.\ref{fig:REW}.

\begin{figure}[t]
	\includegraphics[width=0.5\textwidth]{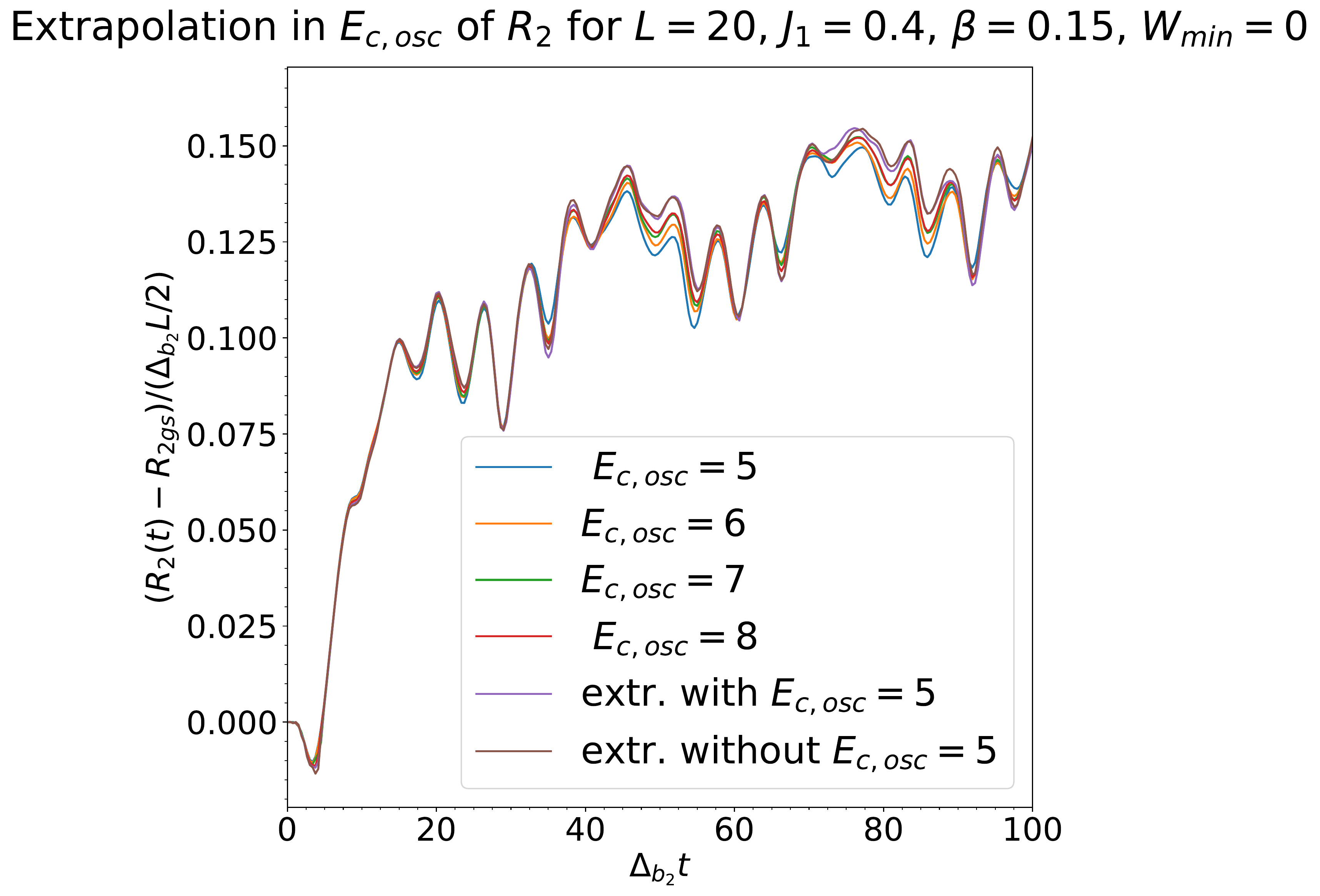}
	\caption{Here we extrapolate the Renyi entropy computed at fixed $L=20,\beta=3/20$ but different $E_{c,osc}$ to $E_{c,osc}=\infty$.} \label{fig:REchi}
\end{figure}

\begin{figure}[t]
	\includegraphics[width=0.5\textwidth]{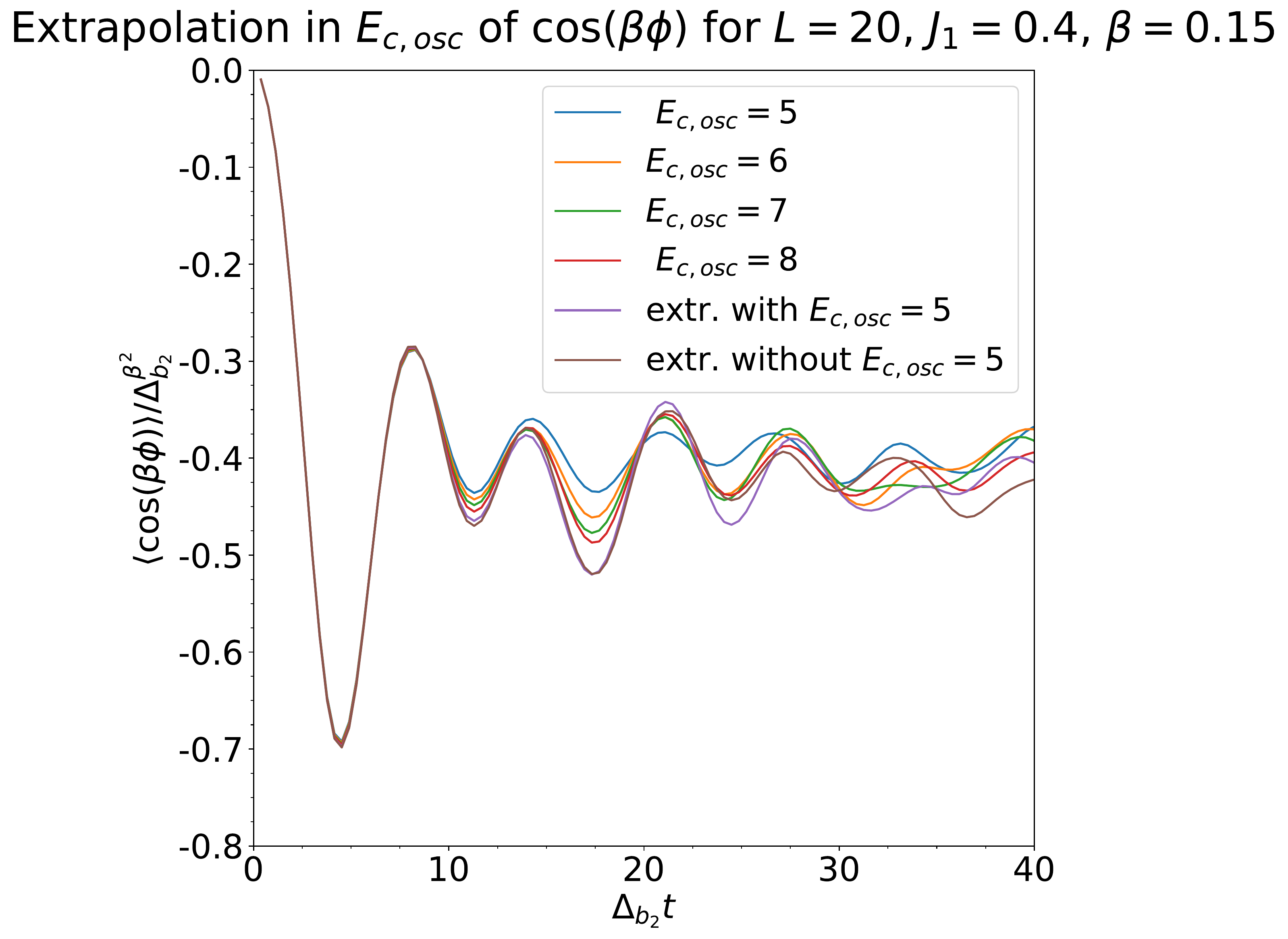}
	\caption{Here we extrapolate the order parameter computed at fixed $\beta=3/30, L=20$ but different $E_{c,osc}$ to $E_{c,osc}=\infty$.} \label{fig:OPchi}
\end{figure}

In order to perform the extrapolation in the cutoff $E_{c,osc}$, we use the already $W_{min}=0$-extrapolated data and the extrapolation form
\begin{equation}
R_{2}(t)-R_{2,gs}=b(t)+c_1(t) E_{c,osc}^{-2-2\beta^2},
\end{equation}
in order to extract $b(t)$, i.e. the second R\'enyi entropies at $E_{c,osc}=\infty$. The same extrapolation form has been used for the cosine operator.  This form is motivated by the known analytic correction to the energy that comes from excluding states from above an energy cutoff, $E_c$ \cite{feverati2008renormalization,watts2012renormalisation,rychkov2015hamiltonian,giokas2011renormalisation,James_2018}.  In Figs. \ref{fig:REchi} and \ref{fig:OPchi}, we show this extrapolations in $E_{c,osc}$, for $R_2(t)$ and $C(t)=\langle\cos(\beta\phi)\rangle(t)$.

We are able to derive error bars from the extrapolation in $E_{c,osc}$.  For all the different cases of $\beta,L$, and $J_1$, we have data for $E_{c,osc}=5,6,7,8$.  We then perform two extrapolations, one using the $E_{c,osc}=5$ data and one not.  The difference between these extrapolations provide an error estimate which is then plotted as error bars in Figs. 1 and 2 of the main text.  We can see that this procedure only produces very small error bars for $R_2(t)$ over a time window of $(0,100\Delta_{b2})$ (see Fig. 1 of main text).  However error bars for the order parameter evolution, $C(t)$ become appreciable after $t>40\Delta_{b2}$.  We thus restrict our presentation of extrapolated $C(t)$ data to this smaller time window.

\subsection{Computation of Power Spectra of $R_2(t)$ and $C(t)$}
Let us explain how the power spectrum presented in Fig. 3 of the main text has been obtained for the set of parameters $R=20, \beta=3/20, J_1=0.4$.
In order to isolate the oscillating behaviour of $R_2(t)$, we did a running time average using
\begin{equation}
\bar{R}_2(t)=\frac{1}{2\Delta_{t.avg}}\sum_{y=t-\Delta_{t.avg}}^{t+\Delta_{t.avg}}R_2(y),
\end{equation}
for $\Delta_{t.avg}=2\pi/\Delta_{b_2}$ over a time window $T\equiv |t_2-t_1|=|132.8-6.8|=126$.   We then performed a discrete Fourier transform (DFT) on the time series $R_2(t)-\bar{R}_2(t)-\varepsilon$, for $t \in [t_1,t_2]$, where $\varepsilon$ was chosen such that $R_2(t)-\bar{R}_2(t)=\varepsilon$ as $t \to \infty$: 
\begin{equation}
\begin{split}\label{eq:ft}
R_{2}(\omega_n)=&\frac{1}{T}\sum_{k=1}^N (R_2(t)-\bar{R}_2(t)-\varepsilon)e^{-i\omega_nk \Delta t}, \quad n=1,\dots, N .\\
\end{split}
\end{equation}
The time averaging serves to suppress frequencies, $\omega_n\ll\omega_{b_2}$.
Here the frequencies, $\omega_n$, of the DFT are defined as $\omega_n=\frac{2\pi }{N \, \Delta t}n$ where  $\Delta t = 0.4$ is the time step and $N=T/\Delta t$. In Fig. 3 of the main text, we plot $|R_2(\omega)|^2$.

The time dependence of $R_2(t)$ can be understood in terms of the eigenstates, $\{|E_i\rangle\}$ of the post-quench Hamiltonian.   A contribution to $R_2(t)$ of the form $|E_i\rangle\langle E_j| E_k\rangle\langle E_l|$ comes with a time dependence, $e^{it(E_i-E_j+E_k-E_l)}$, as explained in Section \ref{sec:tsm}.  So the peaks in the Fourier transform will correspond to quadtuplets $(E_i,E_j,E_k,E_l)$. In general, several possible combinations of the groundstate ($g$) and excited states are present. Looking at the low-lying energies of the states $\ket{E_i}$, we can identify the quadtuplets for each peak in the power spectrum. In the spectrum, the first excited state is the second breather (an excitation involving the first breather alone is forbidden by symmetry) and is denoted as $b_2$. The next two excited states are ($b_1,b_1$) and ($b_1,b_1$)' and are two-particle states of two first breathers.  They are distinguished by the momentum carried by each constituent $b_1$ (although the total momentum of the state sums to zero).  The fourth excited state is the fourth breather, $b_4$.

For the Fourier transform of the cosine operator, we did a DFT on the function $C(t)-C(t=\infty)$, i.e. we subtracted the asymptotic value in order to obtain a power spectrum with $C(\omega=0)=0$.  To perform the DFT, we used as a time window $T=|40-5|=35$.  In this case, the peaks in the Fourier transform correspond to pairs $(E_i,E_j)$ rather than to quadruplets: in the computation, terms like $\langle E_j |\cos(\beta \phi)|E_i\rangle$ appear and provide a $e^{it(E_i-E_j)}$ dependence to $C(t)$. Let us notice that the  dominant peak in the DFT of $C(t)$ is due to the contribution of the first excited state, i.e. the second breather $b_2$, while the amplitude of the next largest peak is due to contributions from $(b_1,b_1)$ and $(b_1,b_1)'$, the two low-lying energy states after $b_2$.

In both DFTs, as with the case of the time series for $R_2(t)$ and $C(t)$, the error bars derive from the use of two sets of extrapolated data, one with $E_{c,osc}=5$ and one without.

\subsection{Presentation of Additional Data for $\beta=1/\sqrt{8}$ and $\beta=1/\sqrt{2}$}
\label{SI:Additional Data}

\begin{figure}[t]
	\includegraphics[width=0.5\textwidth]{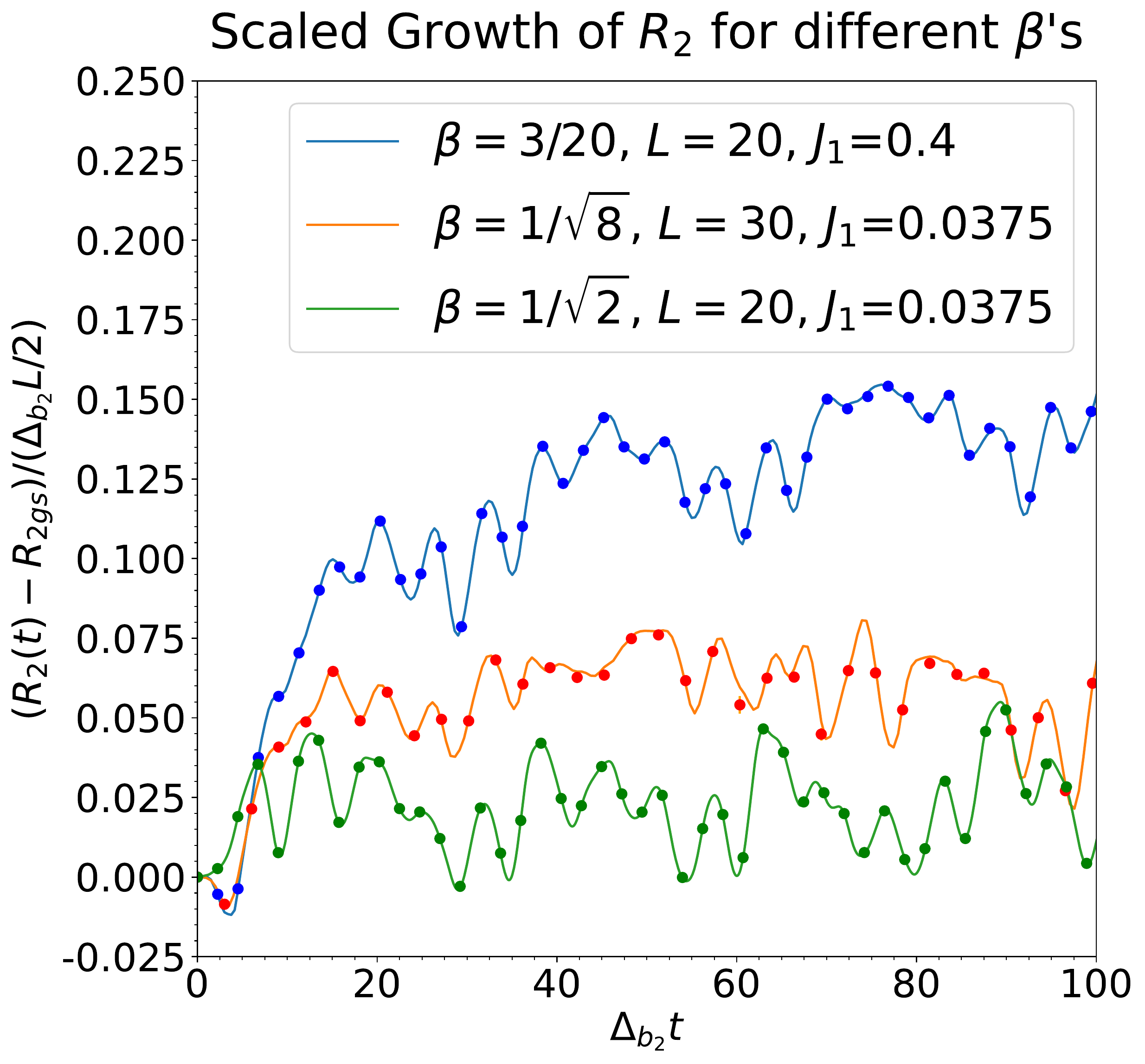}
	\caption{Here is presented the post-quench growth of the second Renyi entropy for three different $\beta$'s. $J_1$ for each is chosen s.t. $\Delta_{b2}L$ is approximately constant for each of the $\beta$'s. The size of the dots represent the error in extrapolating the data in cutoff.} \label{fig:REdiff}
\end{figure}

\begin{figure}[t]
	\includegraphics[width=0.5\textwidth]{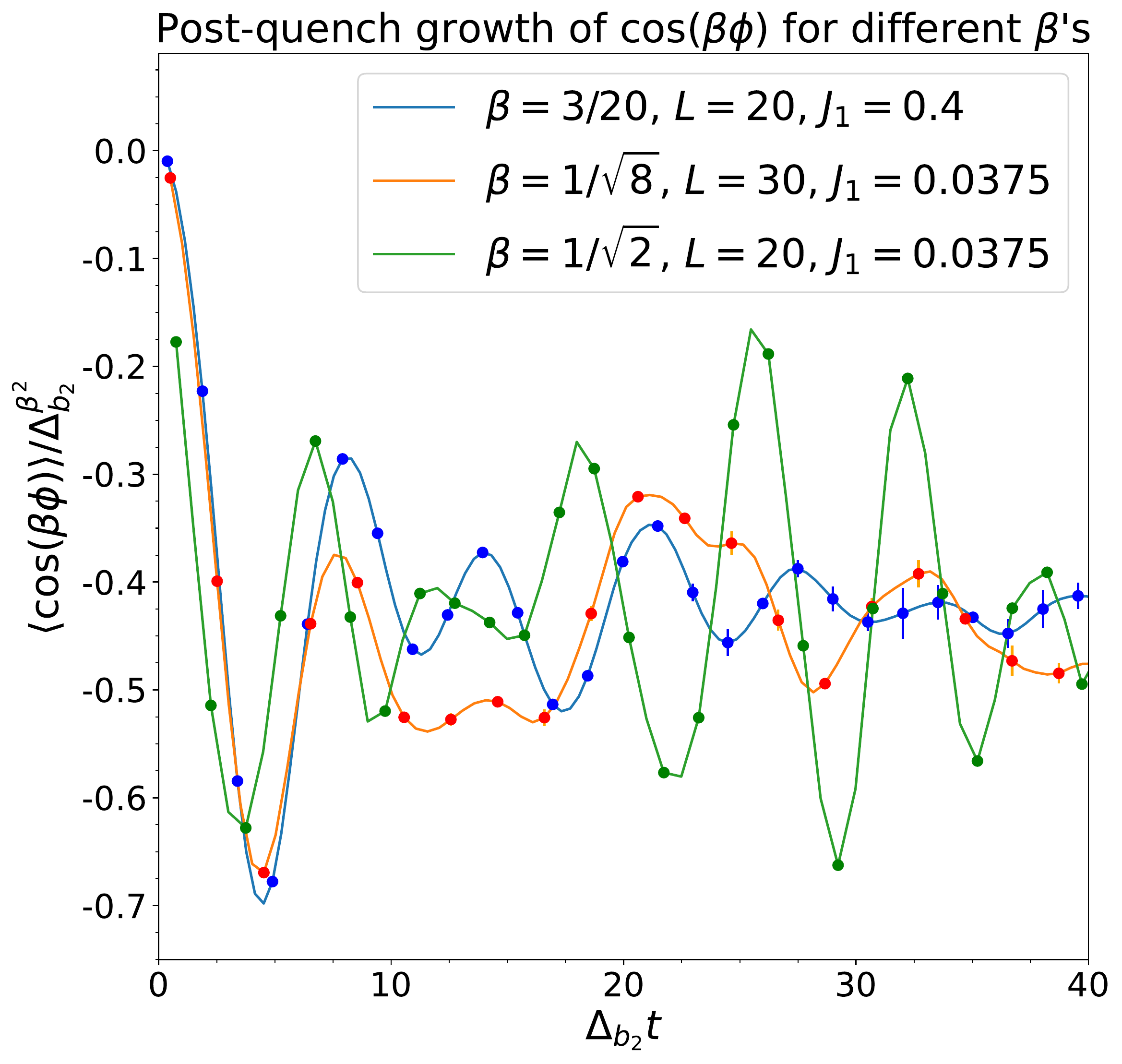}
	\caption{Here is presented the post-quench growth of the order parameter, $\langle\cos(\beta\phi)\rangle$ for three different $\beta$'s. $J_1$ for each is chosen s.t. $\Delta_{b2}L$ is approximately constant. The size of the error bars drawn at each dot represent the error in extrapolating the data in cutoff.} \label{fig:OPdiff}
\end{figure}

In the main text, we presented results mainly for $\beta=3/20$.  Here we also present plots of the time evolution of $R_2(t)$ (see Fig. \ref{fig:REdiff}) and $C(t)$ (see Fig. \ref{fig:OPdiff}) for the values of $\beta=1/\sqrt{8},1/\sqrt{2}$.  We present these data in a way that allows direct comparison of the three values of $\beta$ studied.  We see that as $\beta$ increases, the time needed for $R_2(t)$ to reach its approximate late time asymptote decreases.  We also see that the amplitude of the late time oscillations in both $R_2(t)$ and $C_2(t)$ increase with increasing $\beta$, a likely result of the spectrum of the post-quench Hamiltonian becoming simpler (with far fewer excitation types) and so more discrete.

\begin{figure}[t]
	\includegraphics[width=0.5\textwidth]{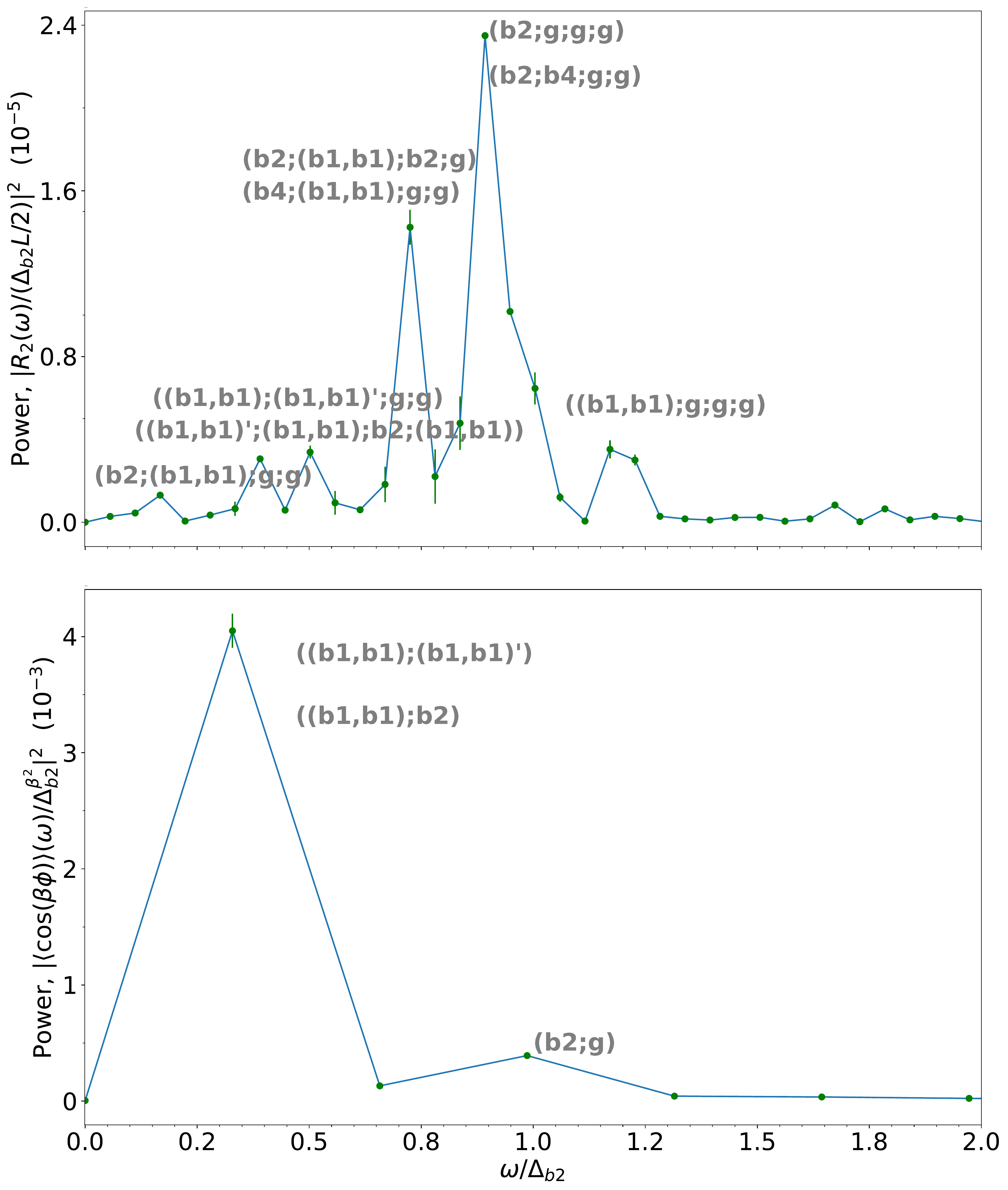}
	\caption{Here for $\beta=1/\sqrt{8}$ is power spectra for the late time oscillations of $R_2$ and $\langle\cos(\beta\phi)\rangle$.} \label{fig:FT353}
\end{figure}

\begin{figure}[t]
	\includegraphics[width=0.5\textwidth]{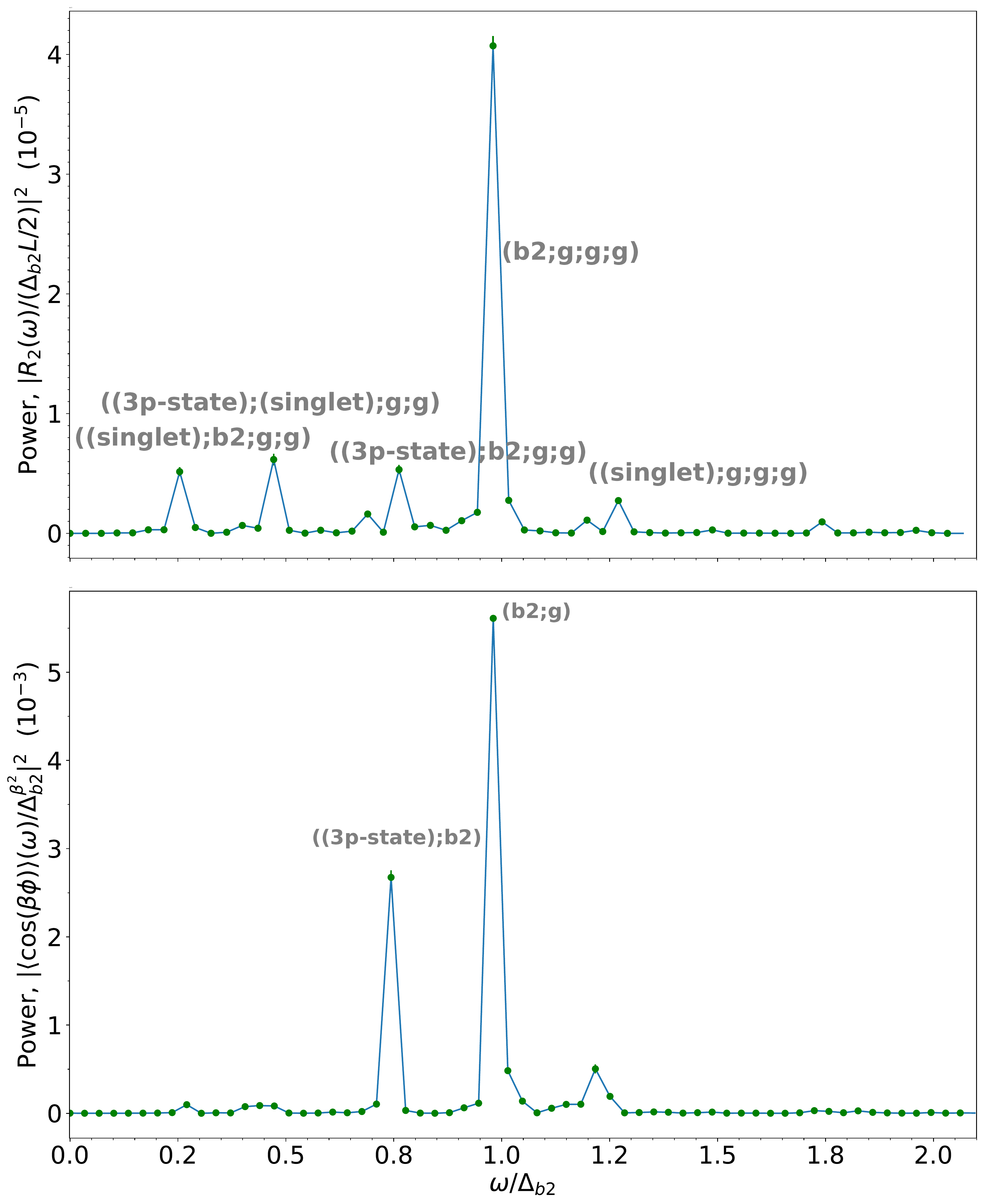}
	\caption{Here for $\beta=1/\sqrt{2}$ is power spectra for the late time oscillations of $R_2$ and $\langle\cos(\beta\phi)\rangle$. } \label{fig:FT707}
\end{figure}

We have also computed the power spectra for the late time oscillations of $R_2(t)$ and $C(t)$ for $\beta=1/\sqrt{8},1/\sqrt{2}$ in Figs. \ref{fig:FT353} and \ref{fig:FT707} respectively. In order to evaluate the power spectrum $R_2(\omega)$ for $\beta=1/\sqrt{8}, J_1=0.0375, L=30$, we repeated the same steps described for $\beta=3/20$, choosing as a time window $T= |t_2-t_1|=|189.8-10.4|$ and time step $a=0.8$ while for the cosine operator we used as a time window $T=|40-10|=30$. Interestingly, the excitation corresponding to the sixth breather ($b_6$) also appears in the power spectrum. \\

For the DFT of $R_2(t)$ for $\beta=1/\sqrt{2}, J_1=0.0375, L=20$, we used as a time window $T= |t_2-t_1|=|192.8-8|$ and time step $a=0.8$.  The spectrum of the model at $\beta=1/\sqrt{2}$ consists of a two breathers, $b_1,b_2$ and two soliton $\pm$.  There is an $SU(2)$ symmetry here and $b_1$ is degenerate in energy with $\pm$ to form an $SU(2)$ triplet while $b_2$ transform as a singlet. In the notation of  Fig. \ref{fig:FT707}, (singlet) refers to two degenerate states corresponding to a linear combination of  $(+,-),(-,+),(b_1,b_1)$, while the $3p$-state refers to a three-particle states composed of $b_1$, the soliton, and the anti-soliton.   

For the power spectrum of $C(t)$, we employed the entire time window of data $T=|200-1|=199$ as our extrapolation procedure was found to be robust at all times.  Thus the peaks of $C(\omega)$ in Fig. \ref{fig:FT707} are much more sharply defined in frequency than for the other cases, $\beta=3/20,1/\sqrt{8}$. \\

\end{document}